\documentstyle[12pt,psfig]{article}

\begin{document}
\bigskip
\bigskip
\bigskip
\bigskip
\begin{center}\bf\large Coulomb Excitation of Double Giant Dipole
  Resonances
\end{center}
\bigskip

\begin{center}
\bigskip
J. Z. Gu and H. A. Weidenm\"uller
\end{center}
\bigskip

\begin{center}
\bigskip
  Max-Planck-Institut f\"ur Kernphysik, Postfach
103980, D-69029 Heidelberg, Germany
\end{center}
\bigskip
\bigskip
\begin{center}
                              Abstract
\end{center}

We implement the Brink--Axel hypothesis for the excitation of the
double giant dipole resonance (DGDR): The background states which
couple to the one--phonon giant dipole resonance are themselves
capable of dipole absorption. These states (and the ones which couple
to the two--phonon resonance) are described in terms of the Gaussian
Orthogonal Ensemble of random matrices. We use second--order
time--dependent perturbation theory and calculate analytically the
ensemble--averaged cross section for excitation of the DGDR. Numerical
calculations illuminate the mechanism and the dependence of the cross
section on the various parameters of the theory, and are specifically 
performed for the reaction $^{208}$Pb + $^{208}$Pb at a projectile 
energy of 640 MeV/nucleon. We show that the contribution of the background
states to the excitation of the DGDR is significant. 
We find that the width of the DGDR, the energy--integrated cross
section and the ratio of this quantity over the energy--integrated
cross section for the single giant dipole resonance, all agree with
experiment within experimental errors. We compare our approach with
that of Carlson {\it et al.} who have used a similar physical picture.
\bigskip


\bigskip
\bigskip
\begin{center}

  e-mail:~~~gu@daniel.mpi-hd.mpg.de
\end{center}
\pagebreak

\section{Introduction}
\label{int}

In peripheral heavy--ion collisions at bombarding energies of several
hundred MeV/nucleon and more, the nuclear giant resonances, in
particular the isovector giant dipole resonance, are excited by
Coulomb excitation. Recently, the excitation of higher dipole modes --
the multi--phonon excitation --  has attracted particular attention.

In the simplest and idealized picture, the giant dipole resonance
(GDR) is a one--phonon state described by the collective motion of
all protons against all neutrons. In addition, there exist also higher
vibrational modes (many--phonon states) that can be excited by
multi--photon absorption. In the harmonic oscillator approximation,
the excitation energy of the $n$--phonon resonance is exactly equal to
$n$ times that of the GDR, and its width is equal to $n$ times the
width of the GDR provided the strength distribution is approximated by
a Lorentzian.

The double giant dipole resonance (DGDR) has been observed in several
nuclei: $^{136}$Xe~\cite{sch93}, $^{197}$Au~\cite{aum93} and $^{208}$
Pb~\cite{rit93, bee93}. Compared to the predictions of the harmonic
picture, the measured cross sections for the DGDR excitation are found
to be enhanced by factors ranging from $1.3$ to $2$ \cite{bau86,
  aum95, eml94}. At the same time the experimentally determined widths
of the DGDR are close to $\sqrt{2}$ times the width of the GDR. These
discrepancies between simple--minded theoretical predictions and
experimental results have attracted much theoretical attention.
Several mechanisms have been studied. We mention anharmonicities of
the collective Hamiltonian \cite{vol95, bor97} and nonlinearities of
the external field \cite{hus99}. Following several earlier papers,
Carlson {\it et al.}~\cite{car991,car992} have recently discussed the
discrepancy using the Brink--Axel hypothesis. This hypothesis states
that a giant resonance is built on top of {\it every} excited nuclear
state \cite{bri62}. Following work by Ko~\cite{ko78}, these authors
considered the contribution to the cross section of DGDR excitation
due to the background states which couple dynamically to the
one--phonon state. It was found that this contribution is sizable.

This result is interesting and calls for further study, especially
since the approach of Carlson {\it et al.}~\cite{car991,car992} uses
approximations which are plausible but not based upon an expansion in
terms of a small parameter, see Section~\ref{comp}. In the present
paper, we apply essentially the same physical picture as Carlson {\it
  et al.} but use a formulation which allows us to derive the DGDR
excitation cross section within perfectly controlled approximations.
The resulting formula is subsequently evaluated numerically. Our
approach makes it possible to clearly identify and calculate the
modification of the DGDR absorption process due to the Brink--Axel
hypothesis.

The paper is organized as follows. We describe our approach in
Section~\ref{hamp}. The background states which couple to the
collective one-- and two--phonon states are complex states. These
states are, therefore, modeled with the help of random--matrix
theory~\cite{meh91, guh98}. In Section~\ref{second}, we derive the
expression for the cross section for DGDR excitation in the framework
of second--order time--dependent perturbation theory. The specific way
in which the Brink--Axel hypothesis has been implemented allows for a
substantial simplification of the resulting expressions
(Section~\ref{ensemble}). An analytical expression for the
ensemble--averaged cross section is derived in Section~\ref{inte}. In
Section~\ref{results}, numerical calculations are used to illuminate
our results and are specifically performed for the reaction $^{208}$Pb
+ $^{208}$Pb at projectile energy 640 MeV/nucleon~\cite{eml94}. We
calculate the width of the DGDR, the energy--integated cross section,
and the ratio of this quantity over the energy--integrated cross
section for single dipole absorption. In Section~\ref{comp}, we
compare our approach and that of Carlson {\it et al.} A brief summary
is given in Section~\ref{sum}.

\section{Hamiltonian of the Projectile}
\label{hamp}

We consider the relativistic Coulomb excitation of the DGDR of the
projectile in a collision with a target.

For the Hamiltonian of the projectile, we use the physical picture
shown in Fig.~1. In the random--phase approximation (RPA), the
one--phonon and the two--phonon states of the projectile are coherent
superpositions of one--particle one--hole (1p1h) states and of
two--particle two--hole (2p2h) states, respectively. The one--phonon
state $|1 0 \rangle$ is the giant dipole mode of the ground state $|0
0 \rangle$, and the two--phonon state $|2 0 \rangle$ is the giant
dipole mode of the one--phonon state. The one--phonon state $|1 0
\rangle$ is dynamically coupled by the nuclear Hamiltonian to more
complex particle--hole configurations $|0 k \rangle$ with
$k=1,\ldots,K$ and $K \gg 1$. This coupling causes the giant dipole
resonance to acquire a spreading width $\Gamma^{\downarrow}_{10}$. The
two--phonon state $|2 0 \rangle$ is likewise coupled to such
configurations. In order to accommodate the Brink--Axel hypothesis, we
group these configurations into two classes. States in the first class
are denoted by $|1 k \rangle$ with $k = 1,\ldots,K$. Each such state
represents the dipole mode of the lower state $|0 k \rangle$. States
in the second class are denoted by $|0 \alpha \rangle$ with $\alpha =
1,\ldots,M$ and $M \gg 1$. These latter states are not coupled by the
dipole operator to lower--lying configurations. The states $|2 0
\rangle$, $|1 k \rangle$ and $|0 \alpha \rangle$ are all dynamically
coupled to each other.

\begin{figure}
\begin{minipage}{23cm}
\centerline{\psfig{figure=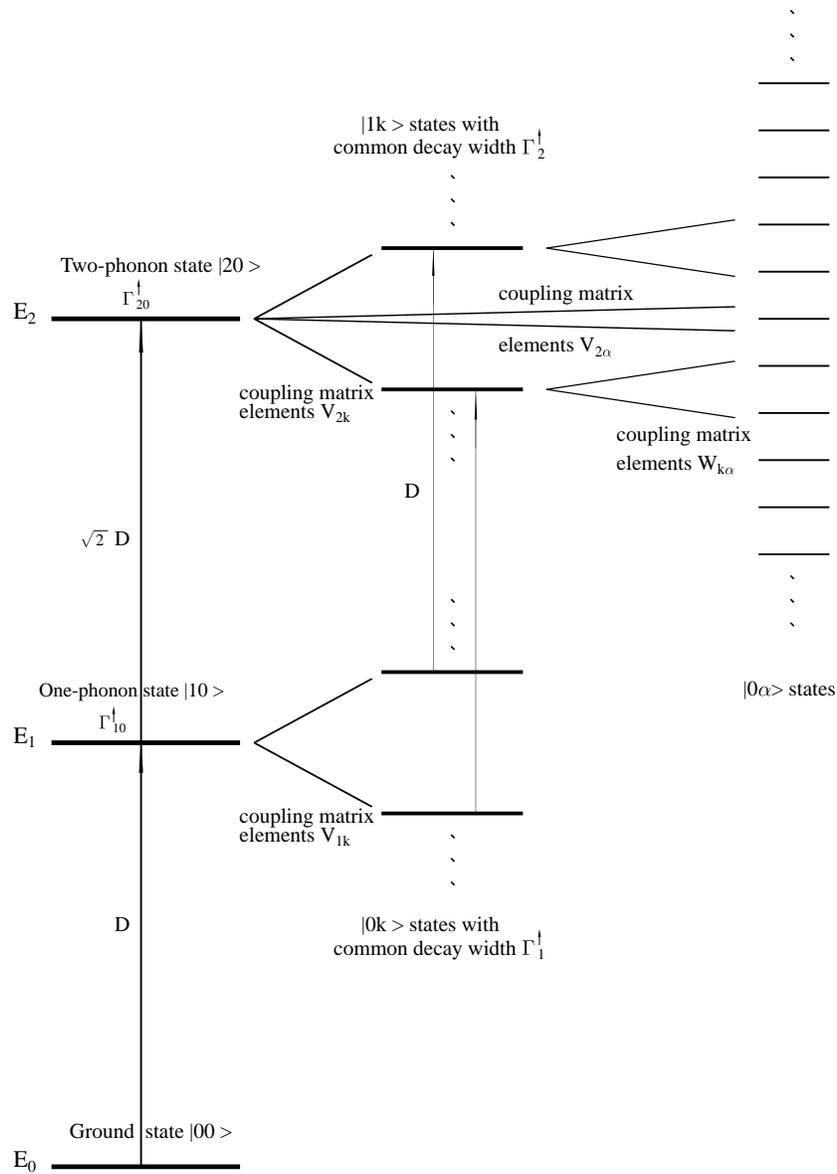,width=22cm,angle=0}}
\end{minipage}
\caption{ Schematic illustration of the DGDR excitation and the
  couplings between the phonon states and many-particle many-hole
  configurations. The level spacings of the configurations are
  exaggerated in the figure. }
\label{fig1}
\end{figure}

According to this picture, the Hamiltonian matrix $H$ has the form
\begin{equation}
\label{ham0}
H =
\left(
\begin{array}{cccccc}
E_{0}&\;\o&\o&\;\o&\o&\o\\
\o\;&E_{1}&V_{1k}&\;\o&\o&\o\\
\o&\;\;V_{l1}&H_{lk}^{(0)}&\;\o&\o&\o\\
\o&\;\o&\o&\;E_{2}&V_{2k^{\prime}}&V_{2\alpha}\\
\o&\;\o&\o&\;\;V_{l^{\prime}2}&H_{l^{\prime}k^{\prime}}^{(1)}&\
W_{l^{\prime} \alpha}\;\;\\ 
\o&\;\o&\o&\;\;V_{\beta 2}&W_{\beta k^{\prime}}&
{\cal H}_{\beta \alpha}^{(0)}\\
%&&&\stackrel{}{---}&\stackrel{}{---}&\stackrel{
%}{---}\\
\end{array}
\right) \ . 
\end{equation}
The matrix $H$ consists of three diagonal blocks. The first block has
dimension one and contains the energy $E_0$ of the ground state $|0 0
\rangle$. The second block has dimension $1 + K$. It contains the
unperturbed energy $E_1$ of the one--phonon state $|1 0 \rangle$, the
elements $H_{k l}^{(0)}$ with $k,l = 1,\ldots,K$ of the Hamiltonian
matrix $H^{(0)}$ governing the states $|0 k \rangle$, and the real
coupling matrix elements $V_{1 k}$ connecting the one--phonon state
with the states $|0 k \rangle$. The third block has dimension $1 + K +
M$. It contains the unperturbed energy $E_2$ of the two--phonon state
$|2 0 \rangle$, the Hamiltonian matrices $H_{k l}^{(1)}$ with $k,l =
1,\ldots,K$ and ${\cal H}_{\alpha \beta}^{(0)}$ with $\alpha,\beta =
1,\ldots,M$ governing the states $|1 k \rangle$ and $|0 \alpha \rangle$,
respectively, and the real coupling matrix elements $V_{2 k}$, $V_{2
  \alpha}$ and $W_{k \alpha}$ connecting the three sets of states. In
this simple picture, we pay no attention to spin and isospin.
Inclusion of these quantum numbers is not difficult but requires a
straightforward extension of our formalism. If, for instance, the
ground state has spin and parity $0^+$, then all states in block two
have spin and parity $1^-$, while the states in block three decay into
three groups having spin/parity values $0^+, 1^+$ and $2^+$,
respectively.

We turn to the statistical assumptions. We describe the $K$ states $|0
k \rangle$ with $K \rightarrow \infty$ in terms of the Gaussian
Orthogonal Ensemble of random matrices (GOE), so that $H^{(0)}$
represents this ensemble. For excitation energies of the giant dipole
resonance (which lie typically between 8 and 15 MeV), this description
seems eminently reasonable, except perhaps for the lightest nuclei.
The spreading width $\Gamma^{\downarrow}_{10} =  2 \pi v^2 / d$ of the
giant dipole resonance is expressed in terms of the mean level spacing
$d$ of the states $|0 k \rangle$ and of the mean square coupling
matrix element $\overline{V_{1k}^2} = v^2$. It was mentioned above
that for each value of $k = 1,\ldots,K$, the state $|1 k \rangle$ is
assumed to be the giant dipole mode of the state $|0 k \rangle$. This
implies that the matrix $H^{(1)}$ is identical to $H^{(0)}$ except
that the center of the semicircle is shifted by $E_2 - E_1$, so that
\begin{equation}
\label{ham1}
H_{k l}^{(1)} = (E_2 - E_1) \delta_{k l} + H_{k l}^{(0)} \ .
\end{equation}
We also assume that $V_{1k} = V_{2k}$. This assumption is not strictly
implied by the Brink--Axel hypothesis because the states $| 1 0
\rangle$ and $| 2 0 \rangle$ (or $| 0 k \rangle$ and $| 1 k \rangle$,
respectively) are not identical. Rather, we have $| 2 0 \rangle
\propto D | 1 0 \rangle$ where $D$ is the dipole operator, and
correspondingly for the states labelled $k$. In practice, the
assumption $V_{1k} = V_{2k}$ implies that the spreading widths for the
couplings $| 1 0 \rangle \Leftrightarrow | 0 k \rangle$ and $| 2 0
\rangle \Leftrightarrow | 1 k \rangle$ are equal. This assumption
seems plausible. We return to this point in Section~\ref{results}. We
note that with these assumptions, the second block of the
matrix~(\ref{ham0}) differs from the corresponding part of the third
block only by $(E_2 - E_1)$ times the unit matrix. The matrix ${\cal
  H}^{(0)}$ is also assumed to represent a GOE, with $M$ taken to
infinity. We assume that ${\cal H}^{(0)}$ and $H^{(0)}$ are uncorrelated.

The states in the second and third block can decay by particle emission.
We take account of this fact by introducing the decay widths
$\Gamma^{\uparrow}_{10}$ and $\Gamma^{\uparrow}_{20}$ of the one--phonon
and the two--phonon states, respectively, and the decay widths
$\Gamma^{\uparrow}_{1}$ and $\Gamma^{\uparrow}_{2}$ of the states $|0 k
\rangle$ and $|1 k \rangle$, respectively. Within the statistical
model, it is obviously adequate to assume that $\Gamma^{\uparrow}_{1}$
and $\Gamma^{\uparrow}_{2}$ are independent of the running index $k
= 1,\ldots,K$. In Section~\ref{ensemble} it is shown that it is not
necessary to assign a decay width to the states $|0 \alpha \rangle$.
The width matrix $\Sigma$ accordingly has the form
\begin{equation}
\label{Sigma}
\Sigma=-(i/2) 
\left(
\begin{array}{cccccc}
0&0&0&0&0&0\\
0&\Gamma^{\uparrow}_{10}&0&0&0&0\\
0&0&\delta_{lk}\Gamma^{\uparrow}_{1}&0&0&0\\
0&0&0&\Gamma^{\uparrow}_{20}&0&0\\
0&0&0&0&\delta_{l^{\prime}k^{\prime}}\Gamma^{\uparrow}_{2}&0\\
0&0&0&0&0&0\\ 
\end{array}
\right). \
\end{equation}
The effective Hamiltonian $H_{\rm eff}$ is then given by
\begin{equation}
\label{eff}
H_{\rm eff} = H + \Sigma \ .
\end{equation}
All four decay widths can be calculated with the help of the optical
model. We return to this point in Section~\ref{results}.

\section{Second--order Time--dependent Perturbation Theory} 
\label{second}

We suppress the intrinsic structure of the target and replace it by a
point source of the electromagnetic field. To describe the DGDR
excitation of the projectile, we use the standard approach to Coulomb
excitation \cite{ald65}: The relative motion of projectile and target
is described classically, and the intrinsic excitation of the
projectile is treated quantum--mechanically.

\subsection{General Approach}
\label{gene}

The time--dependent Schr\"odinger equation
\begin{equation}
\label{Sch}
i\hbar\frac{\partial|\psi(t) \rangle}
{\partial\;t}=(H_{\rm  eff} + H_1(t))|\psi(t)
\rangle
\end{equation}
for the internal wave function $| \psi(t) \rangle$ of the projectile
contains the interaction $H_1(t)$ with the time--dependent
electromagnetic field caused by the relative motion. We consider only
dipole excitation and write $H_1(t) = D h(t)$ as the product of the
dipole operator $D$ and of a time--dependent function $h(t)$. We have
$h(t) \rightarrow 0$ for $t\rightarrow \pm\infty$. A suitable form for
$h(t)$ is given in Section~\ref{results}. The dipole operator $D$
induces transitions between the following pairs of states: $|0 0 
\rangle \rightarrow |1 0 \rangle$, $|1 0 \rangle \rightarrow |2 0
\rangle$, and $|0 k \rangle \rightarrow |1 k \rangle$ for all $k = 1,
\ldots,K$.

We use the interaction representation
\begin{equation}
\label{inter}
|\phi(t) \rangle=\exp(iH_{\rm eff}\:t/\hbar)|\psi(t)
\rangle
\end{equation}
so that 
\begin{equation}  
\label{inter1}  
i\hbar\frac{\partial|\phi(t) \rangle}{\partial\;t}=
\widetilde{H}_{1}(t)|\phi(t) \rangle
\end{equation}
with
\begin{equation}
\label{inter2}
\widetilde{H}_1(t)=\exp(iH_{\rm eff}\:t/\hbar)
H_1(t)\exp(-iH_{\rm eff}\:t/\hbar) \ .
\end{equation}
 
Excitation of the DGDR is a two--step process. We use second--order
time--dependent perturbation theory. With $| \phi(-\infty) \rangle $ =
$|0 0 \rangle$ and $\widetilde{H}_1(t) \rightarrow \;0$ for $t
\rightarrow\pm\infty$, we have
\begin{equation}
\label{inter3}  
| \phi(+\infty) \rangle
= (\frac{1}{i\hbar})^{2}
\int^{\infty}_{-\infty}dt_1\widetilde{H}_1(t_1)\int^{t_1}_{-\infty}
dt_2\widetilde{H}_1(t_2)|0 0 \rangle \ .
\end{equation}
To calculate the intensity $I_2$ for double dipole excitation with
energy transfer $E'$, we first consider the case where
$\Gamma^{\uparrow}_{20}$ and $\Gamma^{\uparrow}_{2}$ both vanish. Let
$|n\rangle$ represent an eigenstate with energy $E_n$ of the submatrix
$H_{{\rm eff}, 33}$ forming the third block of the matrix $H_{\rm
  eff}$. Then $I_2$ is given by
\begin{equation}
\label{inter4}
I_2(E' + E_0) =
\sum_n|\langle\;n|\phi(+\infty)\rangle|^{2}\delta(E'-(E_n - E_0)) \ . 
\end{equation}
For $I_2$ we have used the argument $E' + E_0$ rather than $E'$
because it is convenient to introduce $E = E' + E_0$. Moreover, we use
the identity
\begin{equation}
\label{inter4a}
\sum_n|n\rangle\langle\;n|\delta(E-E_n)=\frac{(-)}{\pi}
{\rm Im}\frac{1}{E^{+}-H_{{\rm eff, 33}}} \ ,
\end{equation}
the definition of $\widetilde{H}_1(t)$ given above, and the fact that
$H_{\rm eff}|0 0 \rangle = E_0|0 0 \rangle$ to rewrite $I_2(E)$ in the
form
\begin{eqnarray}
\label{inter5}
&&I_2(E)=-\frac{1}{\pi\:\hbar^{4}}
\int^{\infty}_{-\infty}dt_1\int^{t_1}_{-\infty}dt_2
\int^{\infty}_{-\infty}dt_{1}' \int^{t_{1}'}_{-\infty}dt_{2}'
\langle 0 0|H_1^{*}(t_{2}')\exp\{-i(H_{\rm eff}^{*}-E_0)(t_{2}' -
t_{1}')/\hbar\}
\nonumber \\
&&\qquad
\times H_1^{*}(t_{1}')
\exp\{-i(H_{\rm eff}^{*}-E_0)t_{1}' / \hbar\}
{\rm Im}(\frac{1}{E^{+}-H_{{\rm eff},33}})
\exp\{i(H_{\rm eff}-E_0)t_1/\hbar\}
\nonumber \\
&&\qquad
\times H_1(t_1)\exp\{i(H_{\rm eff}-E_0)(t_2 - t_1)
/\hbar\}H_1(t_2)|0 0 \rangle \ .
\end{eqnarray}
The form of Eq.~(\ref{inter5}) remains unchanged when we take account
of the finite values of the decay widths $\Gamma^{\uparrow}_{20}$ and
$\Gamma^{\uparrow}_{2}$. Substitution of $H_1(t) = D h(t)$ yields
\begin{eqnarray}
\label{inter6}
&&I_2(E)=-\frac{1}{\pi\:\hbar^{4}}
\int^{\infty}_{-\infty}dt_1\int^{0}_{-\infty}d\tau_2
\int^{\infty}_{-\infty}dt_{1}'\int^{0}_{-\infty}d\tau_{2}'\:
h(t_1)h(\tau_2+t_1)h^{*}(t_{1}')h^{*}(\tau_{2}' + t_{1}')
\nonumber\\
&&\qquad
\times \langle 0 0|D\exp\{-i(H_{{\rm eff},22}^{*} - E_0)\tau_{2}'
/\hbar\}D
\exp\{i(E - E_0)(t_1-t_{1}')/\hbar\}
\nonumber \\
&&\qquad \times {\rm Im}(\frac{1}{E^{+}-H_{{\rm eff},33}})
D\exp\{i(H_{{\rm eff},22} - E_0)\tau_2/\hbar\}
D|0 0 \rangle,
\end{eqnarray}
with $\tau_2 = t_2 - t_1$ and $\tau_{2}' = t_{2}' -t_{1}'$. We have
replaced $\exp\{i(H_{{\rm eff},33}-E_0)(t_1-t_{1}')/\hbar\}$ by
$\exp\{i(E-E_0)(t_1-t_{1}')/\hbar\}$. The remaining two exponentials
in Eq.~(\ref{inter6}) contain the second block $H_{{\rm eff},22}$ of
$H_{\rm eff}$ and can be expressed in terms of Green functions. The
integrand of
\begin{equation}
\label{inter6a}
\int^{\infty}_{-\infty}d\epsilon\frac{1}{\epsilon^{+}-H_{\rm eff}}  
\exp(-i\epsilon\;t/\hbar)
\end{equation}
has poles in the lower half of the $\epsilon$ plane. We may close the
contour of integration in the upper (lower) half plane if $t<$0
($t>$0, respectively). Hence,
\begin{equation}
\label{green2}
\int^{\infty}_{-\infty}d\epsilon\frac{1}{\epsilon^{+}-H_{\rm eff}}
\exp(i\epsilon\;t/\hbar)=-2i\pi\Theta(-t)\exp(iH_{\rm eff}t/\hbar) \ ,
\end{equation}
\begin{equation}
\label{green3}
\int^{\infty}_{-\infty}d\epsilon\frac{1}{\epsilon^{-}-H_{\rm eff}}
\exp(-i\epsilon\;t/\hbar)=2i\pi\Theta(-t)\exp(-iH_{\rm eff}t/\hbar) \ .
\end{equation}
We observe that both the $\tau_2$--integration and the
$\tau_{2}'$--integration are restricted to values $\le$0. We use the
formulas~(\ref{green2}) and ~(\ref{green3}) to find
\begin{eqnarray}
\label{inten1}
&&I_2(E)\;=-\;(\frac{1}{\pi\:\hbar^{4}})
\;\int^{\infty}_{-\infty}\;dt_1\int^{0}_{-\infty}d\tau_2
\int^{\infty}_{-\infty}\;dt_{1}'\;\int^{0}_{-\infty}\;d\tau_{2}'
\;h(t_1)\;h(\tau_2+t_1)\;h^{*}(t_{1}')\;h^{*}(\tau_{2}'+t_{1}')
\nonumber\\
&&\qquad
\times \exp\{i(E-E_0)(t_1-t_{1}')
/\hbar\}
\int^{\infty}_{-\infty}d\epsilon\int^{\infty}_{-\infty}d\epsilon'
\exp\{i(\epsilon-E_0)\tau_2/\hbar\}
\exp\{-i(\epsilon'-E_0)\tau_{2}'/\hbar\}
\nonumber\\
&&\qquad
\times M(E,\epsilon,\epsilon') \ .
\end{eqnarray}
Here, $M$ is a stochastic function of $H_{\rm eff}$ defined by
\begin{equation}  
\label{inten2}
M(E,\epsilon,\epsilon') =
\langle 0 0| D \frac{1}{(\epsilon')^{-}-H_{{\rm eff},22}^{*}} D
{\rm Im}(\frac{1}{E^{+}-H_{{\rm eff},33}}) D
\frac{1}{\epsilon^{+}-H_{{\rm eff},22}} D |0 0\rangle \ .
\end{equation}
To simplify $M$, we observe that $D$ does not connect to the rows and
columns labelled $\alpha$ and $\beta$ in the third block of the matrix
in Eq.~(\ref{ham0}). Therefore, we restrict attention in the third
block to the subspace of states obtained by excluding these rows and
columns. The projection of $H_{{\rm eff},33}$ onto this subspace
(which we denote by $s$) is written as $H_{{\rm eff},s}$, and we have
\begin{equation}
[\frac{1}{E - H_{{\rm eff},33}}]_{s} = \frac{1}{E - H_{{\rm eff},s} -
W^{'}\frac{1}{E^{+}-{\cal H}^{(0)}} (W^{'})^{\dagger}} \ ,
\label{pp1}
\end{equation}
where $W^{'}=(V_{2 \alpha} , W_{l^{'} \alpha})$. We insert this
result back into Eq.~(\ref{inten2}) and find
\begin{equation}
\label{pp3}
M(E,\epsilon,\epsilon') = \langle 0 0 | D \frac{1}{\epsilon' -
  H_{{\rm eff},22}^{*}} D {\rm Im}(\frac{1}{E - H_{{\rm eff},s} - W^{'}
  \frac{1}{E^{+}-{\cal H}^{(0)}} (W^{'})^{\dagger}}) D
  \frac{1}{\epsilon - H_{{\rm eff},22}} D | 0 0 \rangle \ .
\end{equation}
Eqs.~(\ref{inten1}) and (\ref{pp3}) are the central results of this
Section.

\subsection{Simplification of $M$}
\label{ensemble}

The expression for $M$ can be simplified further because $H^{(0)}$ and
${\cal H}^{(0)}$ are uncorrelated. Since ${\cal H}^{(0)}$ appears in
the intensity $I_2(E)$ only via the expression~(\ref{pp3}), we can
perform the average over ${\cal H}^{(0)}$ right away. We have
\begin{equation}
\label{pp2}
\overline{W^{'}\frac{1}{E^{+}-{\cal H}^{(0)}} (W^{'})^{\dagger}} =
- \Sigma^{\downarrow}_s = 
-(i/2)\left( \begin{array}{cc} \Gamma_{20}^{\downarrow}&0\\
0&\delta_{l'k'}\Gamma_{2}^{\downarrow}
\end{array}
\right) \ .
\end{equation}
Here $\Gamma_{20}^{\downarrow}$ is the spreading width of the
two--phonon state due to its coupling to the states $|0 \alpha
\rangle$, and $\Gamma_{2}^{\downarrow}$ is the common spreading width
of the states $|1 k \rangle$ due to their coupling to the states $|0
\alpha \rangle$. We observe that the result of the averaging procedure
is independent of whether or not we include a decay width for the
states $|0 \alpha \rangle$, see the end of Section~\ref{hamp}. We use
the result of Eq.~(\ref{pp2}) in Eq.~(\ref{pp3}). To justify this
step, we observe that a series expansion of the right--hand side of
Eq.~(\ref{pp3}) in powers of the Green function of ${\cal H}^{(0)}$,
followed by averaging over ${\cal H}^{(0)}$, would amount to taking
the average individually over each Green function appearing in the
expansion. This is because the energy argument in all these functions
has a positive imaginary part. Resummation of the result amounts to
using the result of Eq.~(\ref{pp2}) in Eq.~(\ref{pp3}). This yields
for $M$
\begin{equation}  
\label{appr3}
M(E,\epsilon,\epsilon') =
\langle 0 0| D \frac{1}{\epsilon'-H_{{\rm eff},22}^{*}} D {\rm Im}
\bigl( \frac{1}{E- H_{{\rm eff},s} + \Sigma^{\downarrow}_s} \bigr) D
\frac{1}{\epsilon - H_{{\rm eff},22}} D | 0 0 \rangle \ .
\end{equation}
A further simplification of $M$ is possible because the real part of
$H_{{\rm eff},s}$ differs from the real part of $H_{{\rm eff},22}$
only by $(E_2 - E_1)$ times the unit matrix, see Eq.~(\ref{ham1}). To
use this fact, it is necessary to write the matrices appearing in the
three Green functions in Eq.~(\ref{appr3}) more explicitly. We denote
the real part of $H_{{\rm eff},22}$ by $H_{22}$ and have
\begin{equation}
\label{h22}
H_{22} = \left(
\begin{array}{cc}
E_1&V_{1k}\\
V_{l1}&H^{(0)}_{lk}
\end{array}
\right) \ .
\end{equation}
Moreover, we define
\begin{equation}
\label{Sig1}
\Sigma_{1}=(i/2)\left(
\begin{array}{cc}
\Gamma^{\uparrow}_{10}&0\\
0&\delta_{jl}\Gamma^{\uparrow}_{1}
\end{array}
\right) \ ,
\end{equation}
and, with $\widetilde{\Gamma_{20}} = \Gamma^{\uparrow}_{20} +
\Gamma_{20}^{\downarrow}$, $\Gamma_{2} = \Gamma^{\uparrow}_{2} +
\Gamma_{2}^{\downarrow}$,
\begin{equation}
\label{Sig2}
\Sigma_{2}=(i/2)\left(
\begin{array}{cc}
\widetilde{\Gamma_{20}}&0\\
0&\delta_{jl}\Gamma_{2}
\end{array}
\right) \ .
\end{equation}
Then, $M$ takes the form
\begin{equation}
\label{appr4}
M(E,\epsilon,\epsilon') = \langle 0 0| D \frac{1}{\epsilon'-H_{22} -
  \Sigma_1} D {\rm Im}(\frac{1}{E-E_2+E_1-H_{22}+\Sigma_{2}}) D
\frac{1}{\epsilon - H_{22} + \Sigma_1} D | 0 0 \rangle \ .
\end{equation}
A further simplification of this expression is possible with the help
of the Brink--Axel hypothesis and of algebraic identities. In the
spirit of the Brink--Axel hypothesis, we assume that the dipole
transitions $| 0 0 \rangle \rightarrow | 1 0 \rangle$ and $| 0 k
\rangle \rightarrow | 1 k \rangle$ all have identical dipole matrix
elements. We denote all these identical matrix elements also by $D$.
We also take account of the fact that in the harmonic oscillator
picture, the dipole matrix element for the transition $| 1 0 \rangle
\rightarrow | 2 0 \rangle$ is given by $\sqrt{2}D$. Using these facts,
we can express $M$ completely in terms of the Green functions 
\begin{eqnarray}
\label{gr1}
G^{\pm}_{1}(E) = \langle 1 0 |\frac{1}{E - H_{22} \pm \Sigma_{1}} |1 0
  \rangle \ , \nonumber \\  
G^{\pm}_{2}(E) = \langle 1 0 |\frac{1}{E - H_{22} \pm \Sigma_{2}} |1 0
  \rangle \ .
\end{eqnarray}

To this end, we introduce the $(1 + K)$--dimensional unit matrix
${\cal I}$, define $\delta \sigma, \delta \sigma_0, \delta \tau,
\delta \tau_0, \delta \rho$ and $\delta \rho_0$ by
\begin{equation}  
\label{sim1}
\Sigma_2 - \Sigma_{1} = \delta \sigma \; {\cal I} + \delta \sigma_0
\ |1 0 \rangle \langle 1 0|,
\end{equation}
\begin{equation}
\label{sim2}
\Sigma_2 + \Sigma_{1} = \delta \tau \; {\cal I} + \delta \tau_0 \ |1
0 \rangle \langle 1 0|,
\end{equation}
\begin{equation}
\label{sim3}
2\;\Sigma_{1} = \delta \rho \; {\cal I} + \delta \rho_0 \ |1 0
\rangle \langle 1 0| \ ,
\end{equation}
and make use of the identities
\begin{eqnarray}
\label{sim4}
&&\frac{1}{E - H_{22} + \Sigma_2}
\;\frac{1}{\epsilon - H_{22} + \Sigma_1} =
\frac{1}{E - \epsilon + \delta \sigma}(\frac{1}{\epsilon - H_{22} +
  \Sigma_{1}} -\frac{1}{E - H_{22} + \Sigma_{2}})
\nonumber\\
&&\qquad
-\frac{\delta \sigma_0}{E - \epsilon + \delta \sigma}\;
\frac{1}{E - H_{22} + \Sigma_2}|1 0 \rangle \langle 1 0|\frac{1}
{\epsilon - H_{22} + \Sigma_1},
\end{eqnarray}
\begin{eqnarray}
\label{sim5}
&&\frac{1}{E - H_{22} + \Sigma_2}
\;\frac{1}{\epsilon' - H_{22}- \Sigma_1} =
\frac{1}{E - \epsilon' + \delta \tau} (\frac{1}{\epsilon' -
  H_{22} - \Sigma_{1}} - \frac{1}{E - H_{22} + \Sigma_{2}})
\nonumber\\
&&\qquad
-\frac{\delta \tau_0}{E - \epsilon' + \delta \tau}\;
\frac{1}{E - H_{22} + \Sigma_2}|1 0 \rangle \langle 1 0|\frac{1}
{\epsilon' - H_{22} - \Sigma_1} \ ,
\end{eqnarray}
\begin{eqnarray}
\label{sim6}
\frac{1}{\epsilon' - H_{22} - \Sigma_1}
\;\frac{1}{\epsilon - H_{22} + \Sigma_1} =
\frac{1}{\epsilon - \epsilon' + \delta
  \rho}(\frac{1}{\epsilon' - H_{22} - \Sigma_{1}}
-\frac{1}{\epsilon - H_{22} + \Sigma_{1}})
\nonumber\\
\qquad
-\frac{\delta \rho_0}{\epsilon - \epsilon' + \delta \rho}\;
\frac{1}{\epsilon' - H_{22} - \Sigma_1}|1 0 \rangle \langle 1 0|
\frac{1}{\epsilon - H_{22} + \Sigma_1} \ .
\end{eqnarray}
We define $E^{\prime \prime} =  E - E_2 + E_1$ and obtain as a result
\begin{eqnarray}
\label{sim7}
&&M(E,\epsilon,\epsilon') = \frac{i}{2} D^{4} 
\ \biggl(\frac{G^{-}_1(\epsilon')}
{\epsilon - \epsilon' + \delta \rho}(\frac{1}{E^{''} - \epsilon'
  - \delta \sigma} - \frac{1}{E^{''} - \epsilon' + \delta \tau})
\nonumber\\
&&
+ \frac{G^{+}_1(\epsilon)}{\epsilon - \epsilon' + \delta \rho}
(\frac{1}{E^{''} - \epsilon + \delta \sigma} - \frac{1}{E^{''} -
  \epsilon - \delta \tau})
\nonumber\\
&&+(\frac{1}{E^{''} - \epsilon - \delta \tau}\; \frac{1}{E^{''} -
  \epsilon' - \delta \sigma} G^{-}_{2}(E^{''}) - \frac{1}{E^{''} -
  \epsilon + \delta \sigma}\;\frac{1}{E^{''} - \epsilon' + \delta
  \tau} \;G^{+}_{2}(E^{''}))
\nonumber\\
&&+ G^{-}_{1}(\epsilon') G^{+}_{1}(\epsilon)[(\frac{1}{E^{''}-
\epsilon^{'}-\delta \tau}
-\frac{1}{E^{''}-
\epsilon^{'}+\delta \tau})
(\frac{\delta \sigma_{0}}{\epsilon - \epsilon' + \delta \rho}
+\sqrt{2}-1)
\nonumber\\
&&+(\frac{1}{E^{''}-
\epsilon^{'}+\delta \sigma}
-\frac{1}{E^{''}-
\epsilon^{'}-\delta \sigma})
(\frac{\delta \tau_{0}}{\epsilon - \epsilon' + \delta \rho}
+1-\sqrt{2})]
\nonumber\\
&&
- (\frac{\delta \tau_0}{E^{''} - \epsilon - \delta \tau}+\sqrt{2}-1)\;
\frac{G^{-}_{2}(E^{''}) G^{+}_{1}(\epsilon)}
{E^{''} - \epsilon' - \delta \sigma}
- (\frac{\delta \tau_0}{E^{''} - \epsilon^{'}+ \delta \tau}+1-\sqrt{2})\;
\frac{G^{-}_{1}(\epsilon^{'}) G^{+}_{2}(E^{''})}
{E^{''} - \epsilon + \delta \sigma}
\nonumber\\
&&
- (\frac{\delta \sigma_0}{E^{''} - \epsilon' - \delta \sigma}+\sqrt{2}-1)\;
\frac{G^{-}_{2}(E^{''}) G^{-}_{1}(\epsilon^{'})}
{E^{''} - \epsilon' - \delta \tau}
- (\frac{\delta \sigma_0}{E^{''} - \epsilon+ \delta \sigma}+1-\sqrt{2})\;
\frac{G^{+}_{1}(\epsilon) G^{+}_{2}(E^{''})}
{E^{''} - \epsilon + \delta \tau}
\nonumber\\
&&+ G^{-}_{1}(\epsilon')G^{+}_{1}(\epsilon) \ 
\{G^{-}_2(E^{''})[\frac{\delta \tau_0 \delta \sigma_0}
{(E^{''} - \epsilon - \delta \tau)\;
(E^{''} - \epsilon' - \delta\sigma)}
\nonumber\\
&&
+(\sqrt{2}-1)(\frac{\delta \tau_0}{E^{''}-\epsilon-\delta \tau}+\frac{
\delta \sigma_0}{E^{''}-\epsilon'-\delta \sigma})+3-2\sqrt{2}]
\nonumber\\
&&
+G^{+}_2(E^{''})[-\frac{\delta \tau_0 \delta \sigma_0}
{(E^{''} - \epsilon + \delta \sigma)\;
(E^{''} - \epsilon' + \delta\tau)}
\nonumber\\
&&
+(\sqrt{2}-1)(\frac{\delta \sigma_0}{E^{''}-\epsilon+\delta \sigma}+\frac{
\delta \tau_0}{E^{''}-\epsilon'+\delta \tau})+2\sqrt{2}-3]\}
\biggr).
\end{eqnarray}
Terms carrying the factor $(\sqrt{2} - 1)$ arise because not all
dipole matrix elements are equal to $D$. We note that the stochastic
matrix $H_{22}$ appears in Eq.~(\ref{sim7}) only in the Green
functions $G^{\pm}_{1}$ and $G^{\pm}_{2}$.

\section{Ensemble Average and Energy Integration}
\label{inte}

The ensemble average over $H^{(0)}$ affects only the Green functions
$G^{\pm}_{1}$ and $G^{\pm}_{2}$ in Eq.~(\ref{sim7}). The terms in the
first three lines of Eq.~(\ref{sim7}) contain a single Green function
(a ``one--point function'') only. The next four lines contain products
of two Green functions of the type $G^{+}G^{-}$ (a ``two--point
function''), and the last four lines contain two three--point
functions. Calculating the ensemble average of the one--point
functions is straightforward, see Eq.~(\ref{pp2}), and yields
\begin{equation}
\label{op1}
\overline{G^{\pm}_{1}(\epsilon)} = \frac{1}{\epsilon - E_1
\pm \frac{i}{2}\Gamma_{10}} \ ,
\end{equation}
\begin{equation}
\label{op3}
\overline{G^{\pm}_{2}(E)} = \frac{1}{E - E_1
\pm\frac{i}{2}\Gamma_{20}} \ .
\end{equation}
Here $\Gamma_{10}$ = $\Gamma^{\uparrow}_{10} + \Gamma_{10}^{\downarrow}$
and $\Gamma_{20}$ = $\widetilde{\Gamma_{20}} + \Gamma_{10}^{\downarrow}$.

We turn to the ensemble average of the two--point functions. These are
written as the sum of two terms, the product of the averages of the
two Green functions and the average over the product of the
fluctuating parts $\delta G$. For example,
$\overline{G_{1}^{+}(\epsilon) G_{2}^{-}(E^{''})} =
\overline{G^{+}_1(\epsilon)}\;\overline{G^{-}_2(E^{''})} +
\overline{\delta G_{1}^{+}(\epsilon) \delta G_{2}^{-}(E^{''})}$. The
last term can be calculated with the help of the supersymmetry
technique \cite{ver85,gu99}. We do not give the details here because
application of the technique shows that these terms are
negligible. This is because the decay widths $\Gamma^{\uparrow}_{10},
\Gamma^{\uparrow}_{1}, \Gamma^{\uparrow}_{20}, \Gamma^{\uparrow}_{2}$
turn out to be orders of magnitude larger than the mean level spacing
$d$ of the states $|0 k \rangle$, see Section~\ref{results}. This fact
leads to an exponential suppression of the fluctuating part. As a
result we have, to a very good approximation,
\begin{equation}
\label{tp3}
\overline{G_{1}^{+}(\epsilon)\; G_{2}^{-}(E'')} =
\overline{G_{1}^{+}(\epsilon)} \;\;\overline{G_{2}^{-}(E'')} \ .
\end{equation}
We have already remarked that the average of the product of two
retarded Green functions (or of two advanced Green functions) is
trivially equal to the product of the averages. Thus, we find that the
averages over all two--point functions in Eq.~(\ref{sim7}) are equal
to the product of the averages of the two Green functions. We turn to
the average over the three--point functions. Here, a similar argument
is expected to apply. Unfortunately, the supersymmetry technique is
not capable of giving an analytical result in this case, and we have
used numerical simulations to estimate the fluctuating part. For the
nucleus $^{208}$Pb, for instance, we have found that the contribution
of the fluctuating part is less than one part in $10^{4}$. Hence, we
have
\begin{equation}
\label{appr1}
\overline{G_{1}^{+} G_{1}^{-} G_{2}^{\pm}} \cong \overline{G_{1}^{+}
  G_{1}^{-}} \,\,\ \overline{G_{2}^{\pm}} = \overline{G_{1}^{+}} \,\,\
  \overline{G_{1}^{-}} \,\,\  \overline{G_{2}^{\pm}} \ ,
\end{equation}
and the ensemble average of $M(E,\epsilon,\epsilon')$ and $I_2(E)$
can be computed in terms of the ensemble--averaged one--point functions.

\subsection{Total Intensity}

Using contour integration, we calculate the integrals over the
variables $\epsilon$ and $\epsilon'$. This calculation is lengthy
but straightforward. Using Eq.~(\ref{sim7}), we find 
\begin{eqnarray}
\label{main}
&&\overline{I_2(E)} = 2 \pi\,i(\frac{D}{\hbar})^4
\int_{-\infty}^{+\infty} dt_1
\int_{-\infty}^{0} d\tau_2 \int_{-\infty}^{+\infty} dt_{1}'
\int_{-\infty}^{0} d\tau_{2}' \nonumber \\
&&\qquad
h(t_1) h(t_1 + \tau_2) h^{*}(t_{1}') h^{*}(t_{1}' + \tau_{2}')
\exp[i(E-E_0)(t_1-t_{1}') / \hbar]
\nonumber\\
&&\qquad
\times \biggl( f_0\Theta(\tau_2-\tau_{2}')
\exp[-i( E_1-E_0 + \frac{i}{2} \Gamma_{10} ) \tau_{2}' / \hbar]
\exp[ i( E_1-E_0 + \frac{i}{2} \Gamma_{10} - i \Delta\rho ) \tau_2/
\hbar]
\nonumber\\
&&\qquad \qquad
\times (f_1^{+}(E) - f_2^{-}(E))
\nonumber\\
&&\qquad
+ f_0 \Theta(\tau_2 - \tau_{2}')
\exp[-i(E^{''}-E_0 + i \Delta \tau) \tau_{2}' / \hbar]
\exp[ i(E^{''}-E_0 + i \Delta \sigma) \tau_2 / \hbar]
(f_2^{+}(E) - f_1^{+}(E))
\nonumber\\
&&\qquad
- f_0 \Theta(\tau_{2}' - \tau_2)
\exp[-i(E^{''}-E_0 - i \Delta \sigma) \tau_{2}' / \hbar]
\exp[ i(E^{''}-E_0 - i\Delta \tau) \tau_2 / \hbar]
(f_2^{-}(E) - f_1^{-}(E))
\nonumber\\
&&\qquad
- f_0 \Theta(\tau_{2}' - \tau_2)
\exp[-i(E_1-E_0 - \frac{i}{2} \Gamma_{10} + i \Delta \rho) \tau_{2}'
/ \hbar] \exp[ i(E_1-E_0 - \frac{i}{2} \Gamma_{10}) \tau_2 / \hbar]
\nonumber\\
&&\qquad \qquad \times (f_1^{-}(E) - f_2^{+}(E))
\nonumber\\
&&\qquad +(\sqrt{2}-1+i \Delta \sigma_0 f_2^{-}(E))
(- \overline{G_2^{-}(E^{''})} 
\nonumber \\ 
&&\qquad + (1+i\Delta \tau_0 \overline{G_2^{-}(E^{''})})
f_1^{-}(E)) \exp[-i(E_1-E_0 + \frac{i}{2} \Gamma_{10}) \tau_{2}' /
\hbar] \exp[ i(E^{''}-E_0 - i\Delta \tau) \tau_2 / \hbar]
\nonumber\\
&&\qquad +(1-\sqrt{2}+i \Delta \sigma_0 f_2^{+}(E))
(- \overline{G_2^{+}(E^{''})} \nonumber \\ 
&&\qquad + (1 - i \Delta \tau_0 \overline{G_2^{+}(E^{''})})
f_1^{+}(E)) \exp[-i(E^{''}-E_0 + \Delta \tau) \tau_{2}' / \hbar]
\exp[i(E_1-E_0 - \frac{i}{2} \Gamma_{10}) \tau_2 / \hbar]
\nonumber\\
&&\qquad -[(1-\sqrt{2}+i\Delta \sigma_0\;f^{+}_{2}(E))
(1-i \Delta \tau_0 \overline{G_2^{+}(E^{''})})
f_1^{+}(E) 
\nonumber\\
&&\qquad + (\sqrt{2}-1+i \Delta \sigma_0\;f^{-}_{2}(E)
)(1 + i \Delta \tau_0
\overline{G_2^{-}(E^{''})}) f_1^{-}(E)) 
\nonumber\\
&&\qquad + \Delta \rho_0 \frac{1}{\Gamma_{10} - \Delta
  \rho}(f_2^{-}(E) - f_2^{+}(E)) + (\sqrt{2} - 1)
(\overline{G^{-}_{2}(E^{''})} - \overline{G^{+}_{2}(E^{''})})]
\nonumber\\
&&\qquad \qquad \times \exp[-i(E_1-E_0 + \frac{i}{2}
\Gamma_{10}) \tau_{2}' / \hbar] \exp[i(E_1-E_0 - \frac{i}{2}
\Gamma_{10}) \tau_2 / \hbar] \biggr) \ ,
\end{eqnarray}
with
\begin{eqnarray}
\label{ffs}
&&f_1^{\pm}(E) = \frac{1}{E - E_2 \mp \frac{i}{2} \Gamma_{10} \pm i
  \Delta \tau} \ ,
\nonumber \\
&&f_2^{\pm}(E) = \frac{1}{E - E_2 \pm \frac{i}{2} \Gamma_{10} \pm i
  \Delta \sigma} \ ,
\nonumber \\
&&\qquad f_0 = 1 - \frac{\Delta \rho_0}{\Gamma_{10} - \Delta \rho} \ .
\end{eqnarray}
Eq.~(\ref{main}) constitutes the main result of the theoretical part
of this paper. Under perfectly controlled approximations, we have
derived an expression for the intensity which embodies the Brink--Axel
hypothesis and which describes the formation of the DGDR as a
transport process.

\subsection{Intraband Intensity}

We now calculate the intensity $\overline{I_2^{\rm intra}(E)}$ which
would result if only the ground state and the one--phonon state
could absorb dipole radiation. In other words, we suppress dipole
absorption by the states labelled $|0 k \rangle$, although we do keep
the dynamical coupling of all the states as described by the matrix
$H_{\rm eff}$ defined in Eq.~(\ref{eff}). We do so in order to
distinguish the dipole excitation taken into account by the usual
approach to the problem, from the transport process described by
Eq.~(\ref{main}). This intensity is defined as
\begin{eqnarray}
\label{inten7}
&&I^{\rm intra}_2(E) =
- \frac{2}{\pi \hbar^{4}}\int_{-\infty}^{+\infty}dt_1
\int_{-\infty}^{0}d\tau_2\int_{-\infty}^{+\infty}dt_{1}'
\int_{-\infty}^{0}d\tau_{2}'
\ h(t_1)h(t_1+\tau_2)h^{*}(t_{1}')h^{*}(t_{1}' + \tau_{2}')
\nonumber\\
&&\qquad
\times \exp[i(E-E_0)(t_1-t_{1}')/\hbar]
\int^{\infty}_{-\infty}d\epsilon\int^{\infty}_{-\infty}d\epsilon'
\exp\{i(\epsilon-E_0)\tau_2/\hbar\}
\nonumber\\
&&\qquad
\times \exp\{-i(\epsilon' - E_0)\tau_{2}' / \hbar\}
\langle 0 0|D|1 0 \rangle \langle 1 0|\frac{1}{\epsilon' - H_{22} -
  \Sigma_1} |1 0 \rangle \langle 1 0|D|2 0 \rangle
\nonumber\\
&&\qquad \times \langle 2 0 |{\rm Im}(\frac{1}{E - E_2 + E_1 - H_{22}
  + \Sigma_2})|2 0 \rangle 
\nonumber \\
&&\qquad \times \langle 2 0|D|1 0 \rangle 
\langle 1 0|\frac{1}{\epsilon-H_{22}+\Sigma_1}|1 0
\rangle \langle 1 0|D|0 0 \rangle \ .
\end{eqnarray}
The ensemble average is given by
\begin{eqnarray}
\label{inten8}
&&\overline{I_2^{\rm intra}(E)}
=2 \pi \times 2 \times
(\frac{D}{\hbar})^{4}\int_{-\infty}^{+\infty}dt_1
\int_{-\infty}^{0}d\tau_2\int_{-\infty}^{+\infty}dt_{1}'
\int_{-\infty}^{0}d\tau_{2}'
\ h(t_1)h(t_1+\tau_2)h^{*}(t_{1}')h^{*}(t_{1}' + \tau_{2}')
\nonumber\\
&&\qquad
\times \exp[i(E-E_0)(t_1-t_{1}') / \hbar]
\frac{\Gamma_{20}}{(E-E_2)^{2}+\frac{1}{4}\Gamma_{20}^{2}}
\exp\{i(E_1-E_0-i/2\Gamma_{10})\tau_2/\hbar\}
\nonumber\\
&&\qquad
\times \exp\{-i(E_1-E_0+i/2\Gamma_{10})\tau_{2}' / \hbar\} \ .
\end{eqnarray}
The factor $2$ arises because the dipole transition $| 1 0 \rangle
\rightarrow | 2 0 \rangle$ is twice as strong as the transition $| 0 0
\rangle \rightarrow | 1 0 \rangle$. We note that when $\Gamma_{10}$
and $\Gamma_{20}$ go to zero, $\overline{I^{\rm intra}_{2}(E)}$ is
identical to the harmonic limit denoted by $I^{\rm har}_{2}(E)$. 

For the discussion of our results and the comparison with experimental
data, it is useful to also give the average intensity
$\overline{I_1(E)}$ for the excitation of the GDR. It is calculated
similarly and given by
\begin{eqnarray}
\label{inten9}
\overline{I_1(E)}
=\frac{D^{2}}{2\pi \hbar^{2}}\int_{-\infty}^{+\infty}dt_1
\int_{-\infty}^{+\infty}dt_{1}'
\ h(t_1)h^{*}(t_{1}')
\exp[i(E-E_0)(t_1-t_{1}') / \hbar]
\frac{\Gamma_{10}}{(E-E_1)^{2}+\frac{1}{4}\Gamma_{10}^{2}} \ .
\end{eqnarray}

\subsection{Cross Sections}

The cross section for the DGDR excitation is given by
\begin{equation}
\label{cross1}
\sigma_{2}(E)=2\pi\int^{\infty}_{b_{\rm
    min}}b\;db\;\overline{I_2(E)} \ .  
\end{equation}
Here $b_{\rm min}$ is the minimal impact parameter which is introduced
in order to account for the strong absorption that occurs as soon as
the colliding nuclei come within reach of their nuclear forces. The
cross section $\sigma_1(E)$ for single dipole excitation is defined
analogously. We also define
\begin{equation}
\label{sum2}
\sigma_{2}(E)^{\rm intra} =
2\pi\int^{\infty}_{b_{\rm min}}b\;db\;\overline{I_2^{\rm intra} (E)}
\end{equation}
and
\begin{equation}
\label{en1}
\sigma_{2}^{\rm har}(E) = 2 \pi
\int_{b_{\rm min}}^{\infty}b\:db\;I^{\rm har}_{2}(E) \ .
\end{equation}
We define two enhancement factors. The first one compares our result
with the harmonic limit and is defined as
\begin{equation}
\label{en2}
R_1 = \frac{\int_{-\infty}^{\infty} \ dE \sigma_{2}(E)}
{\int_{-\infty}^{\infty} \ dE \sigma_{2}^{\rm har}(E)} \ .
\end{equation}
The second factor compares our result with the cross section
calculated without the Brink--Axel hypothesis and is defined as
\begin{equation}
\label{en3}
R_2 = \frac{\int_{-\infty}^{\infty} \ dE \sigma_{2}(E)}
{\int_{-\infty}^{\infty} \ dE \sigma_{2}^{\rm intra}(E)} \ .
\end{equation}
It is also of interest to compare $\sigma_{2}^{\rm intra}(E)$ with the
harmonic approximation. This is accomplished by the third factor
\begin{equation}
\label{en4}
R_1^{\rm intra} = \frac{\int_{-\infty}^{\infty} \ dE \sigma_{2}^{\rm
    intra}(E)} {\int_{-\infty}^{\infty} \ dE \sigma_{2}^{\rm har}(E)}
    \ .
\end{equation}
The contribution of the extraband excitation is measured by the
following two ratios,
\begin{equation}
\label{en4a}
R_1^{\rm extra} = \frac{\int_{-\infty}^{\infty} \ dE
  (\sigma_{2}(E)-\sigma_{2}^{\rm intra}(E))} {\int_{-\infty}^{\infty}
  \ dE \sigma_{2}^{\rm har}(E)} \ ,
\end{equation}
\begin{equation}
\label{en4b}
R_2^{\rm extra} = \frac{\int_{-\infty}^{\infty} \ dE
  (\sigma_{2}(E)-\sigma_{2}^{\rm intra}(E))}{\int_{-\infty}^{\infty} \
  dE \sigma_{2}^{\rm intra}(E)} \ .
\end{equation}
We note that $R_{1}^{\rm extra}$ and $R_{2}^{\rm extra}$ also account
for the contribution of interference terms between the intraband and
extraband excitations.

\section{Numerical Results}
\label{results}

For the calculation of the decay widths $\Gamma^{\uparrow}_{10},
\Gamma^{\uparrow}_{1}, \Gamma^{\uparrow}_{20}$ and
$\Gamma^{\uparrow}_{2}$, we have used the computer code developed by
E. Sheldon and V. C. Rogers \cite{she73}. It contains a global
optical--model potential to compute the transmission coefficients of
nucleons. These in turn were used to determine the mean absorption
cross section using the exciton model~\cite{man76}: The average decay
rate of a state with n excitons into a channel where the nucleon has
(asymptotic) kinetic energy $\epsilon_k$ is given by \cite{man76}
\begin{equation}
\label{exciton1}
w_n(\epsilon_k) = \frac{2m \epsilon_k}{\pi^{2} \hbar^{3}}
\sigma(\epsilon_k) \frac{\rho_{n-1}(p-1,h,U-E_b - \epsilon_k)}
{\rho_n(p,h,U)} \ ,
\end{equation}
where m is the reduced mass, U the excitation energy of the residual
nucleus, $E_b$ the separation energy of particle b and
$\sigma(\epsilon_k)$ the mean absorption cross section. The function
$\rho_n(p,h,U)$ is the $n$--exciton state density with excitation
energy U,
\begin{equation}
\label{exciton3}
\rho_n(p,h,U)=\frac{g}{p!h!(p+h-1)!}(Ug)^{p+h-1} \ .
\end{equation}
Here $g$ is single--particle level density, $p$ and $h$ are the numbers
of particles and holes, and $n = p+h$. The decay width is given by
\begin{equation}
\label{exciton2}
\Gamma^{\uparrow} = \hbar \sum_b \int^{E-E_b}_{0}w_n(\epsilon_k) {\rm
  d} \epsilon_k \ .
\end{equation}
The decay widths $\Gamma_{10}^{\uparrow}$ and $\Gamma_{1}^{\uparrow}$
are evaluated at energy $E_1$, and $\Gamma_{20}^{\uparrow}$ and
$\Gamma_{2}^{\uparrow}$ at energy $E_2$.

In the long--wavelength approximation~\cite{ber94} where the impact
parameter $b$ is large compared to the nuclear radius $r$, the
time--dependent function $h(t)$ is given by
\begin{equation}
\label{dipole1}
h(t) = \frac{\gamma}{b^{2}} \frac{1 + \gamma\tau} {(1 +
  \tau^{2})^{3/2}} - i \ \frac{\gamma v \omega}{b} \frac{1}{(1 +
  \tau^{2})^{1/2}} \ .
\end{equation}
Here $\tau=\gamma\;v\;t/b$ with $\gamma = 1 / \sqrt{(1-v^{2}/c^{2})}$,
$v$ is the relative velocity, and $\omega=(E_f-E_i)/\hbar$. The
validity of the long--wavelength approximation was
checked~\cite{lan97} for giant resonance excitation in the process
$^{208}$Pb \ + \ $^{208}$Pb at 640 MeV/A and was found to be accurate
to within a few percent.

We consider the DGDR excitation of a $^{208}$Pb projectile, incident
on a $^{208}$Pb target. This reaction has been studied at 640 MeV/A at
the GSI/SIS, Darmstadt~\cite{eml94}. We apply the formalism developed
in the preceding Sections to calculate the cross section, enhancement
factors, and the DGDR width. Using the method described above and with
$p = h$ the values $p = 1$ for $\Gamma_{10}^{\uparrow}$, $p = 2$ for
$\Gamma_{1}^{\uparrow}$, $p = 2$ for $\Gamma_{20}^{\uparrow}$, and $p
= 3$ for $\Gamma_{2}^{\uparrow}$, we find for the decay widths the
values $\Gamma_{10}^{\uparrow} = 0.11$ MeV, $\Gamma_{20}^{\uparrow} =
0.026$ MeV, $\Gamma_{1}^{\uparrow} = 0.30$ MeV and
$\Gamma_{2}^{\uparrow} = 0.16$ MeV. We set $E_1 - E_0$ and $E_2 - E_0$
equal to their experimental values 13.5 MeV and 27.0 MeV, respectively.
The parameter $\omega$ appearing in the parametrization~(\ref{dipole1})
of $h(t)$ was accordingly chosen as $\hbar \omega = 13.5$ MeV. Various
parametrizations of $b_{min}$ with regard to the nuclear system have
been proposed. Most widely used are those of Ref.~\cite{ben89},
\begin{equation}
\label{b1}
b_{min} = 1.34 (A_{1}^{1/3} + A_{2}^{1/3} - 0.75 (A_1^{-1/3} +
A_{2}^{-1/3})) \ ({\rm fm})
\end{equation}
and of Ref.~\cite{kox87}, 
\begin{equation}  
\label{b2}
b_{min} = 1.1(A_{1}^{1/3} + A_{2}^{1/3} + \frac{A_1^{1/3} A_2^{1/3}}
{A_1^{1/3} + A_{2}^{1/3}} - 1.9) \ ({\rm fm}) \ .
\end{equation}
For the system $^{208}$Pb \ + \ $^{208}$Pb, Eqs.~(\ref{b1}) and
~(\ref{b2}) yield $b_{min} =  15.7$ fm and $b_{min} = 14.4$ fm,
respectively. We have used the mean of these estimated values,
$b_{min} =  15.05$ fm.

In order to calculate the absolute values of the cross sections
we evalute the dipole matrix elements by means of the sum rule for
dipole transition
\begin{equation}
\label{dsum1}
\sum_{f}\omega_{fi}|D_{fi}^{(m)}|^{2}=\frac{3e^{2}}{8\pi
  m_{N}}\frac{NZ}{A}=S_{D} \ . 
\end{equation}
Here $D_{fi}^{(m)}=\int rY_{1m}(\hat{r})\rho(r)dv$.
In practice, we use
\begin{equation}
\label{dsum2}
\sum_{f}\omega_{fi}|D^{2}|=\frac{4\pi}{3}S_{D},
\end{equation}
with $D=\int x\rho(r)dv=\int z\rho(r)dv$. Unless otherwise stated, we
always use 100\% of the sum rule.

Particular attention must be paid to the spreading widths as these
determine largely the outcome of the calculation. Here, we repeat a
statement already made in the Introduction: This paper is primarily
devoted to the implementation of the Brink--Axel hypothesis, to the
derivation of an expression for the average cross section based on
this hypothesis and on controlled approximations, and to the
investigation of the consequences of this hypothesis for the
excitation cross section of the DGDR. In this sense, we consider the
various spreading widths as parameters. We are not primarily concerned
with assigning physically realistic values to the
$\Gamma^{\downarrow}$'s. Still, it is worthwhile to comment on some
problems that arise if one tries to do so.

The parameters are the spreading width $\Gamma_{2}^{\downarrow}$ of
the states $| 1 k \rangle$, the spreading width
$\Gamma_{10}^{\downarrow}$ of the one--phonon state $| 1 0 \rangle$
(by assumption this width also describes the mixing of the
two--phonon state $| 2 0 \rangle$ with the states $| 1 k \rangle$),
and the spreading width $\Gamma_{20}^{\downarrow}$ which describes the
mixing of the two--phonon state $| 2 0 \rangle$ with the states $| 0
\alpha \rangle$. The total spreading width of the two--phonon state is
accordingly given by $\Gamma^{\downarrow}_{{\rm total} \ 2 0} =
\Gamma_{10}^{\downarrow} + \Gamma_{20}^{\downarrow}$. We know
experimentally the spreading width $\Gamma_{10} = 4.0$
MeV~\cite{eml94} of the one--phonon state. Theoretically $\Gamma_{10}$
is given by $\Gamma_{10} = \Gamma_{10}^{\uparrow} +
\Gamma_{10}^{\downarrow}$. Typically, three effects contribute to the
observed spreading width of the GDR: The coupling to the complex modes
of excitation contained in our Hamiltonian of Eq.~(\ref{ham0}), the
spread in energy of the single--particle modes which contribute to the
GDR (Landau damping), and the coupling to low--lying vibrational modes
\cite{ber83}. This makes it difficult to estimate the spreading width
$\Gamma^{\downarrow}_{{\rm total} \ 2 0}$ of the two--phonon state.
Neglecting the contribution of Landau damping and of the vibrational
modes to $\Gamma_{10}^{\downarrow}$ altogether and using the exciton
model in the form of Ref.~\cite{her92} (we identify the $n$--phonon
state as a $n$--particle $n$--hole configuration), we find that the
spreading width $\Gamma^{\downarrow}_{{\rm total} \ n 0}$ of the
$n$--phonon states is proportional to $2 n \ {\rm Im} \ [V_{\rm
  opt}(E_n / 2 n)]$. Here $V_{\rm opt}(E)$ is the depth of the 
optical--model potential at energy $E$, and $E_n$ is the excitation
energy. This relation derives from the fact that each exciton
(particle or hole) carries the average energy $(E_n / 2 n)$. We
observe that for $n = 1$ and $n = 2$, the energy arguments of $V_{\rm
  opt}(E_n / 2 n)$ coincide, yielding $\Gamma^{\downarrow}_{{\rm
    total} \ 2 0} = 2 \Gamma^{\downarrow}_{1 0}$. This result is in
keeping with the predictions of the harmonic picture for the
$n$--phonon states and gives $\Gamma_{20}^{\downarrow} =
\Gamma_{10}^{\downarrow} = 4.0$ MeV. Comparison with experimental
data will show that this is an overestimate which we ascribe to the
fact that Landau damping and coupling to the vibrational modes
contribute differently to $\Gamma_{10}^{\downarrow}$ and to
$\Gamma_{20}^{\downarrow}$. It turns out that
$\Gamma_{20}^{\downarrow} \sim 0.5 \ \Gamma_{10}^{\downarrow}$ yields
results which are in good agreement with the data. Similarly, using
the measured value for $\Gamma_{10}^{\downarrow}$ and the exciton
model to estimate $\Gamma_{2}^{\downarrow}$ yields an unreasonably
large value. We often use $\Gamma_{2}^{\downarrow} = 0.5$ MeV and
remark only that our results are insensitive to this choice, see
Fig.~5.

Prior to our giving detailed results for the reaction $^{208}$Pb on
$^{208}$Pb at 640 MeV/A (the case studied experimentally), we discuss
the dependence of our results on some of the relevant parameters. The
following calculations were all done for the reaction $^{208}$Pb on
$^{208}$Pb as described above except that the energy, the decay widths
and the spreading widths were varied.

We investigate the dependence of the cross sections for the GDR 
and DGDR excitation on the spreading width $\Gamma_{10}^{\downarrow}$
and on the projectile energy $E_{p}/A$. It is instructive to begin
with the GDR excitation. Fig.~2 shows the energy--integrated cross
section $\sigma_1$ for the GDR excitation as a function of the
spreading with $\Gamma_{10}^{\downarrow}$ at various projectile
energies. The dependence of $\sigma_1$ on $\Gamma_{10}^{\downarrow}$
changes significantly with projectile energy. For the lower projectile
energies, the cross section first increases strongly then more slowly
and finally decreases as $\Gamma_{10}^{\downarrow}$ increases. For
large $E_{p}/A$ (10000 MeV) the cross section decreases monotonically
with increasing $\Gamma_{10}^{\downarrow}$. This dependence can be
understood qualitatively with the help of the Weizsaecker-Williams
(WW) method of equivalent photons \cite{eml94,ber88}. In this method
the transition probability for the GDR excitation at impact parameter
b is given by
\begin{equation}
\label{photon1}
P_{\rm GDR}(b) = \int N(b,E_{\gamma}) \sigma_{\rm phot}(E_{\gamma})
\frac{dE_{\gamma}}{E_{\gamma}} \ .
\end{equation}
Here $N(b,E_{\gamma})$ denotes the number of equivalent photons of
frequency $E_{\gamma}/\hbar$ at impact parameter $b$, a function of
projectile energy, while $\sigma_{\rm phot}(E_{\gamma})$ is the
photoabsorption cross section which is well described by a Lorentzian.
For small values of the spreading width $\Gamma_{10}^{\downarrow}$,
only the photons with frequency close to the central frequency
$(E_{1}-E_{0})/\hbar$ substantially contribute to the transition
probability. As $\Gamma_{10}^{\downarrow}$ increases, an ever wider
band of photons contributes to and enhances $P_{\rm GDR}(b)$. On the
other hand, the widening of the normalized Lorentzian causes the
absorption strength of the photons with frequencies around the central
frequency to decrease, which suppresses $P_{\rm GDR}(b)$. These
effects compete against each other. The result depends on the
distribution of available photons over the relevant frequency range.
For small values of $E_{p}/A$, $N(b,E_{\gamma})$ decreases rapidly
with increasing $E_{\gamma}$ \cite{llo90}. In this case the gain of
$P_{\rm GDR}(b)$ exceeds the loss unless $\Gamma_{10}^{\downarrow}$
becomes large. As a result, the GDR cross section (an integral of
$P_{\rm GDR}(b)$ over impact parameter) increases with
$\Gamma_{10}^{\downarrow}$ up to a certain value of the spreading
width beyond which the loss surpasses the gain. At very high
projectile energy the curve of $N(b,E_{\gamma})$ versus $E_{\gamma}$
becomes flat \cite{llo90}. The loss always exceeds the gain, and the
cross section decreases monotonically with $\Gamma_{10}^{\downarrow}$.
Since $N(b,E_{\gamma})$ increases with $E_{p}/A$, so do both $P_{\rm
  GDR}(b)$ and cross section. Fig.~2 shows that the enhancement of the
cross section due to the spreading width is most pronounced when
$E_{p}/A$ is around several hundred MeV. We refer to this enhancement
as to the damping enhancement. Taking for projectile energy and
spreading width the experimental values $E_{p}/A$ = 640 MeV and
$\Gamma_{10}^{\downarrow}$ = 4.0 MeV, we find for the ratio of the
energy--integrated cross section with damping to that without damping
the value 1.20.

\begin{figure}
\begin{minipage}{18cm}
\centerline{\psfig{figure=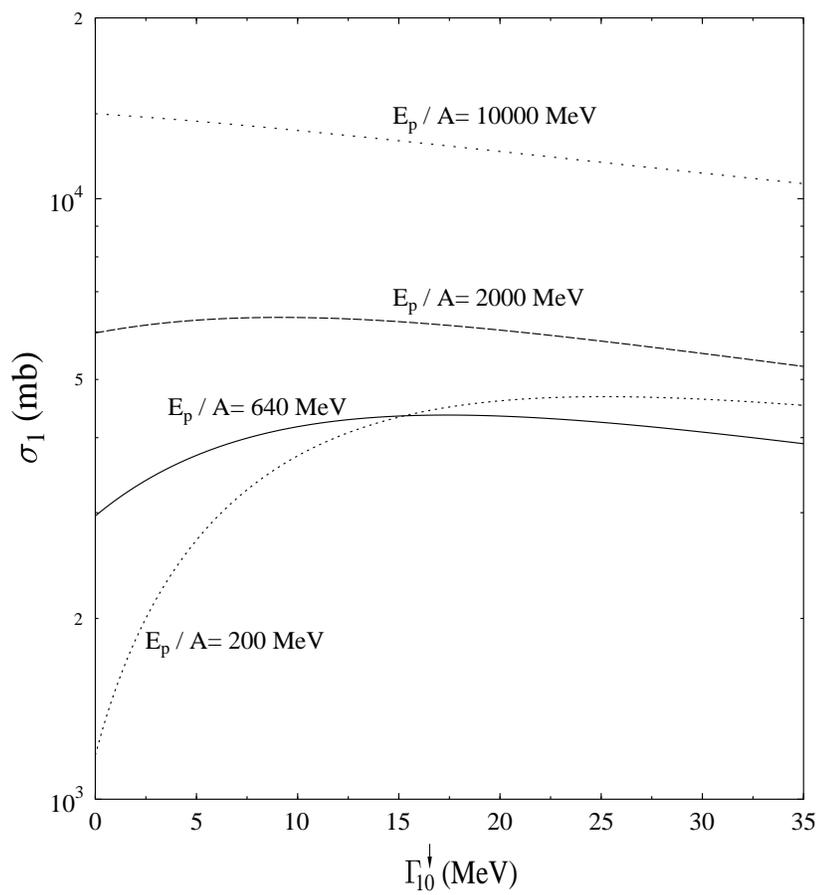,width=15cm,angle=0}}
\end{minipage}
\caption{Energy--integrated cross section for the GDR excitation as a 
function of spreading width $\Gamma_{10}^{\downarrow}$ at different 
projectile energies $E_p/A$, with $\Gamma_{10}^{\uparrow}$ = 0.11
MeV.}
\label{fig2}
\end{figure}

Fig.~3 shows for several projectile energies the dependence of the
energy--integrated cross section for the DGDR excitation on the
spreading width $\Gamma_{10}^{\downarrow}$. We use the five ratios
defined in Section 4.3 to measure the cross sections. We keep the
decay widths and the ratio $\Gamma_{10}^{\downarrow} /
\Gamma_{20}^{\downarrow}$ fixed. We focus first on the intraband cross
section measured by $R^{\rm intra}_{1}$. The dependence of the
intraband cross section on spreading width and projectile energy is
similar to that of $\sigma_{1}$ (Fig.~2): At small projectile
energies, $R^{\rm intra}_{1}$ first increases then decreases with
increasing $\Gamma_{10}^{\downarrow}$, while it decreases
monotonically with $\Gamma_{10}^{\downarrow}$ at high energies. This
is because the intraband excitation is governed by two transition
probabilities: one for the transition $| 0 0 \rangle \rightarrow | 1 0
\rangle$, the other for the transition $| 1 0 \rangle \rightarrow | 2
0 \rangle$. Each of the two probabilities can be described by
Eq.~(\ref {photon1}). Thus, the intraband cross section depends on
spreading width and projectile energy in a similar way as $\sigma_{1}$.
Some of the probability flux feeding the one--phonon state $| 1 0
\rangle$ flows into the states $| 0 k \rangle$, and -- in the
intraband approximation -- is not available for the second dipole
excitation. This flow increases with $\Gamma_{10}^{\downarrow}$.
Therefore, the enhancement of $R^{\rm intra}_{1}$ only survives for
rather low projectile energies, and the range in which $R^{\rm intra}$
increases with the spreading width is significantly reduced compared
to that of $\sigma_1$. The extraband excitation measured by
$R_{1}^{\rm extra}$ (compared with the harmonic limit) and $R_{2}^{\rm
  extra}$ (compared with the intraband cross section) increases
monotonically with $\Gamma_{10}^{\downarrow}$. For small projectile
energy ($E_{p}/A$ = 200 MeV) and small values of
$\Gamma_{10}^{\downarrow}$ both ratios have small negative values.
However, in most cases the extraband excitation gives a positive
contribution to the total cross section.

\begin{figure}
\begin{minipage}{18cm}
\centerline{\psfig{figure=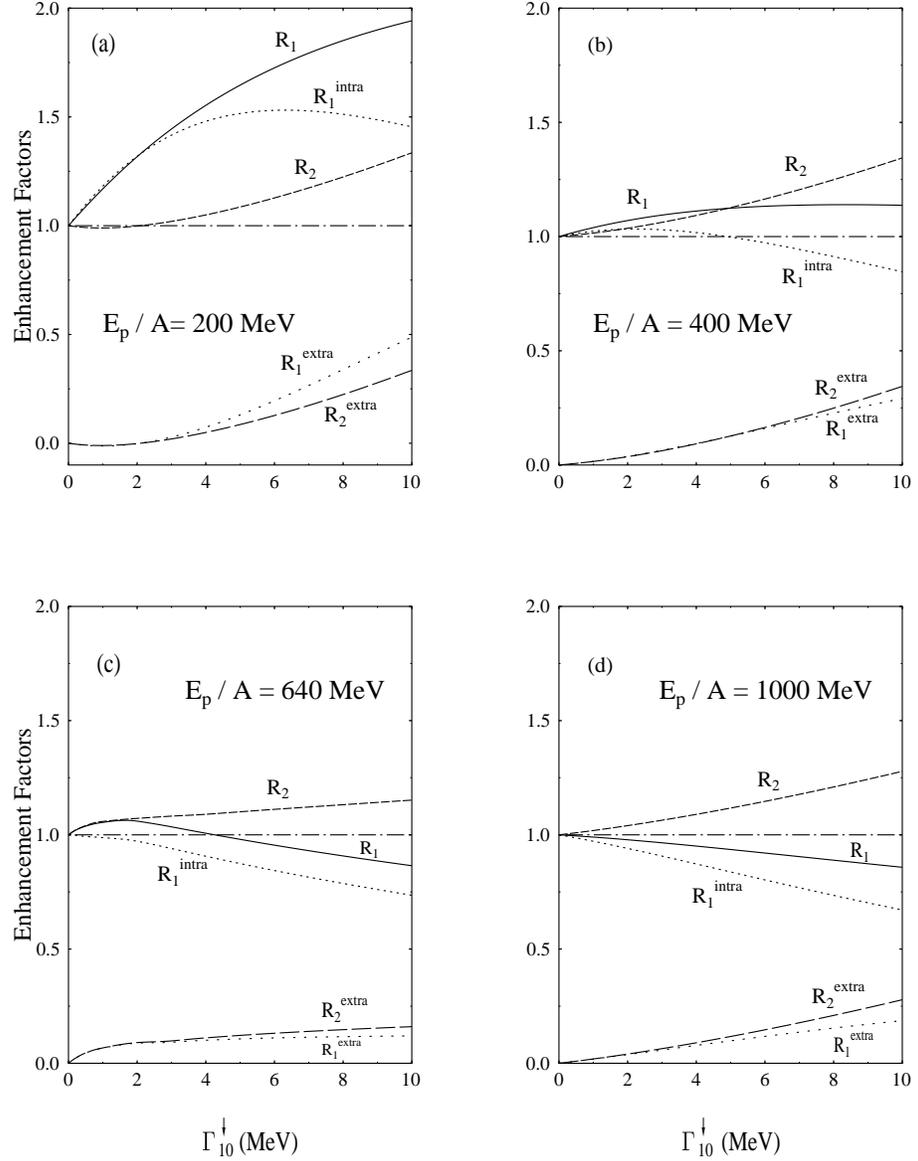,width=16cm,angle=0}}
\end{minipage}
\caption{Enhancement factors for the energy--integrated cross sections
  for DGDR excitation as described in the text as functions of
  $\Gamma_{10}^{\downarrow}$ at different projectile energies. 
  Parameter values are
  $\Gamma_{10}^{\uparrow}$ = 0.11, $\Gamma_{20}^{\uparrow}$ = 0.026,
  $\Gamma_{1}^{\uparrow}$ = 0.30, $\Gamma_{2}^{\uparrow}$ = 0.16,
  $\Gamma_{20}^{\downarrow} = 0.5 \Gamma_{10}^{\downarrow}$ and
  $\Gamma_{2}^{\downarrow} = 0.5$ (MeV).}
\label{fig3}
\end{figure}

The total cross section is measured by $R_{1}$ (compared with the
harmonic limit) and by $R_{2}$ (compared with the intraband cross
section). The ratio $R_{2}$ increases monotonically with
$\Gamma_{10}^{\downarrow}$ and is larger than unity except for small
projectile energy and small $\Gamma_{10}^{\downarrow}$. At the smaller
projectile energies, $R_{1}$ first increases and later decreases with
$\Gamma_{10}^{\downarrow}$. The turning point shifts towards larger
values of $\Gamma_{10}^{\downarrow}$ as $E_p/A$ decreases. With
increasing $E_{p}/A$ the range where $R_{1}$ is larger than unity
disappears. An enhancement of the total cross section compared with
the harmonic limit can, therefore, be expected only for low projectile
energy. Moreover, this enhancement is not mainly due to the extraband
excitation ($R_{1}^{\rm extra}$ and $R_{2}^{\rm extra}$ being small),
but is mainly due to the damping enhancement discussed for $\sigma_1$
above. For the case of $E_{p}/A$ = 200 MeV and
$\Gamma_{10}^{\downarrow}$ = 4.0 MeV, $R_{1}$ reaches the value
1.50. For $E_{p}/A$ = 640 MeV where the measurement was performed and
$\Gamma_{10}^{\downarrow}$ = 4.0 MeV, we find $R_{1}$ = 1.01: The gain
of the total cross section due to extraband transitions just
compensates the loss due to the reduction of the intraband cross
section, resulting in a zero net enhancement for the total cross
section compared with the harmonic limit.
  
Fig.~4 shows the dependence of the cross sections and enhancement
factors for the DGDR excitation on projectile energy. In Fig.~4(a),
$\sigma_{2}$, $\sigma_{2}^{\rm intra}$, $\sigma_{2}^{\rm har}$ and
$\sigma_{2}^{\rm extra}$ are the total, intraband, harmonic limit and
extraband energy--integrated cross sections, respectively. (Actually,
$\sigma_{2}^{\rm extra}$ is not a cross section itself but rather the
difference of two cross sections). All of the cross sections increase
monotonically with $E_{p}/A$. The total cross section largely stems
from intraband excitation. The extraband transition contributes not
more than 10\% compared with the harmonic limit. For $E_{p}/A <$ 420
MeV, $\sigma_{2}^{\rm intra}$ is larger than $\sigma_{2}^{\rm har}$.
This indicates once again the damping enhancement which occurs at
lower projectile energies where the extraband contribution is
negligible. For very low projectile energy, the enhancement factors
$R_{1}$ and $R_{1}^{\rm intra}$ can reach several hundred percent. This
results from the damping enhancement. When $E_{p}/A$ is larger than
500 MeV, $R_{2}$, $R_{2}^{\rm extra}$ and $R_{1}^{\rm extra}$ are
constant with values below 1.10 and 0.1, respectively.

\begin{figure}
\begin{minipage}{15cm}
\centerline{\psfig{figure=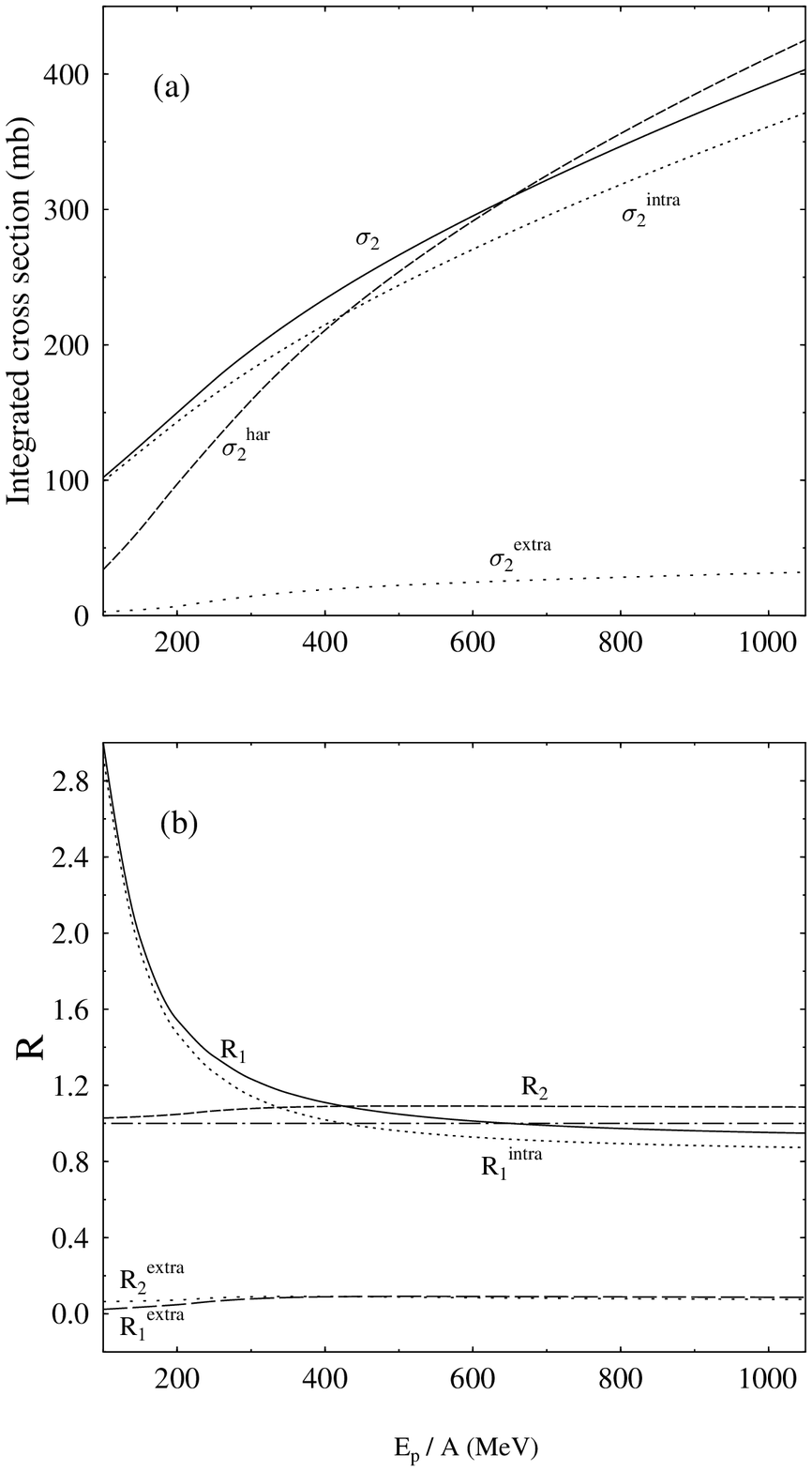,width=13cm,angle=0}}
\end{minipage}
\caption{(a) Energy--integrated cross sections and (b) enhancement
factors for the DGDR excitation versus $E_{p}/A$. Parameter values are
$\Gamma_{10}^{\uparrow}$ = 0.11, $\Gamma_{20}^{\uparrow}$ = 0.026,
$\Gamma_{10}^{\downarrow}$ = 4.0, $\Gamma_{1}^{\uparrow}$ = 0.30,
$\Gamma_{2}^{\uparrow}$ = 0.16, $\Gamma_{20}^{\downarrow} =
0.5 \Gamma_{10}^{\downarrow}$ and $\Gamma_{2}^{\downarrow} = 0.5$
(MeV).}
\label{fig4}
\end{figure}

We turn to the dependence of the cross section for the DGDR excitation
on the parameters $\Gamma_{10}^{\uparrow}$, $\Gamma_{20}^{\uparrow}$,
$\Gamma_{1}^{\uparrow}$, $\Gamma_{2}^{\uparrow}$,
$\Gamma_{20}^{\downarrow}$ and $\Gamma_{2}^{\downarrow}$, using
$R_{1}$ and $R_{1}^{\rm intra}$. We note that the parameters
$\Gamma_{2}^{\uparrow}$, $\Gamma_{2}^{\downarrow}$,
$\Gamma_{20}^{\uparrow}$ and $\Gamma_{20}^{\downarrow}$ appear in the
expression for the cross section only in the combinations
$\Gamma_{2}^{\uparrow}+\Gamma_{2}^{\downarrow}$ and
$\Gamma_{20}^{\uparrow}+\Gamma_{20}^{\downarrow}$.

Fig.~5(a) shows $R_1$ and $R_1^{\rm intra}$ as functions of
$\Gamma_{10}^{\uparrow}$. The ratios decrease with increasing
$\Gamma_{10}^{\uparrow}$, and are reduced by almost a factor of two
when $\Gamma_{10}^{\uparrow}$ is as large as several MeV. For
realistic values of $\Gamma_{10}^{\uparrow}$ the reduction is quite
small, however. An increase of $\Gamma_{10}^{\uparrow}$ reduces the
lifetime of the one--phonon state and suppresses the intraband
excitation which dominates the DGDR excitation (Fig.~4). Both ratios
depend only weakly or not at all on $\Gamma_{1}^{\uparrow}$
(Fig.~5(b)). It is clear that the dominant intraband excitation is
independent of $\Gamma_{1}^{\uparrow}$. The slow decrease of $R_1$
with increasing $\Gamma_{1}^{\uparrow}$ is due to the loss of
extraband transition strength. However, the latter gives only a small
contribution to $R_1$. In Fig.~5(c) and Fig.~5(d) we display the
dependence of $R_1$ and of $R_1^{\rm intra}$ on $\Gamma_{2}^{\uparrow}
+ \Gamma_{2}^{\downarrow}$ and on $\Gamma_{20}^{\uparrow} +
\Gamma_{20}^{\downarrow}$, respectively. In both cases $R_1$ increases
with the increase of the variable. However, this increase is so slow
that the precise choice of these parameters is irrelevant for our
calculations. The two variable combinations on the abscissas of
Figs.~5(c) and 5(d) do not cause any loss of probability flux feeding
the DGDR. Therefore, we attribute the slow rise of $R_1$ in both
figures to damping enhancement. This is obvious for Fig.~5(c) where
the intraband cross section is independent of $\Gamma_{2}^{\uparrow} +
\Gamma_{2}^{\downarrow}$ and where the increase of $R_1$ is due to the
damping enhancement for the transition $| 0 k \rangle \rightarrow | 1
k \rangle$. In Fig.~5(d) damping enhancement works also for $R_1^{\rm
  intra}$. It is interesting to see that the damping enhancement
survives (albeit weakly) even if $E_{p}/A$ is as high as 640 MeV.

\begin{figure}
\begin{minipage}{19cm}
\centerline{\psfig{figure=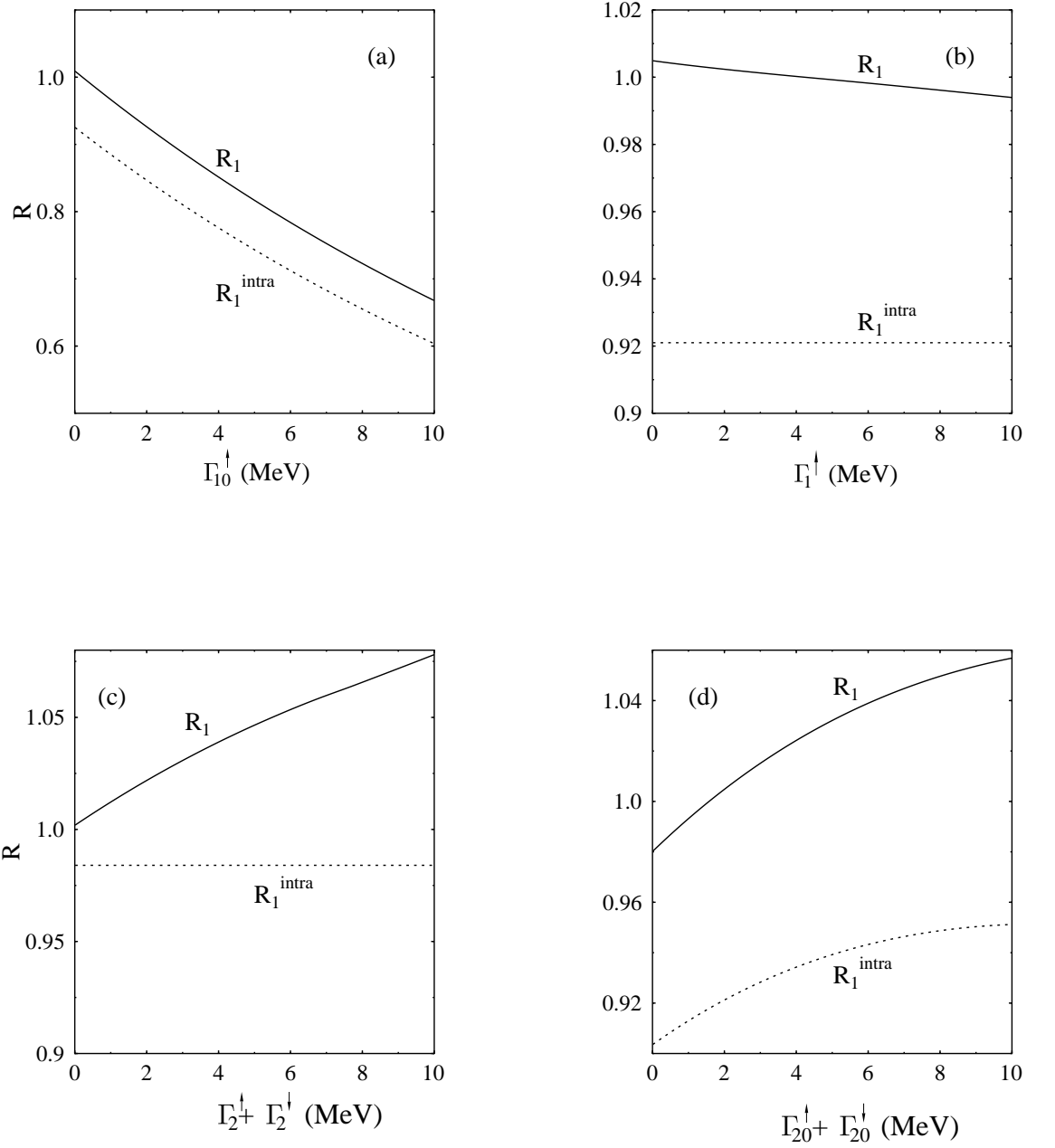,width=18cm,angle=0}}
\end{minipage}
\caption{Enhancement factors of the energy--integrated cross sections
  for DGDR excitation versus (a) $\Gamma_{10}^{\uparrow}$, (b)
  $\Gamma_{1}^{\uparrow}$, (c) $\Gamma_{2}^{\uparrow} +
  \Gamma_{2}^{\downarrow}$ and (d) $\Gamma_{20}^{\uparrow} +
  \Gamma_{20}^{\downarrow}$. Parameter values are $E_{p}/A$ = 640,
  $\Gamma_{10}^{\downarrow}$ = 4.0,
  $\Gamma_{20}^{\downarrow} = 0.5 \Gamma_{10}^{\downarrow}$, 
  (a) $\Gamma_{20}^{\uparrow}$ = 0.026, 
  $\Gamma_{1}^{\uparrow}$ = 0.30, $\Gamma_{2}^{\uparrow}$ = 0.16,
  $\Gamma_{2}^{\downarrow}=0.5$; 
  (b) $\Gamma_{10}^{\uparrow}$ = 0.11, $\Gamma_{20}^{\uparrow}$ =
  0.026, $\Gamma_{2}^{\uparrow}$ = 0.16, $\Gamma_{2}^{\downarrow} = 0.5$;
  (c) $\Gamma_{10}^{\uparrow}$ = 0.11, $\Gamma_{20}^{\uparrow}$ = 0.026, 
  $\Gamma_{1}^{\uparrow}$ = 0.30; 
  (d) $\Gamma_{10}^{\uparrow}$ = 0.11, 
  $\Gamma_{1}^{\uparrow}$ = 0.30, $\Gamma_{2}^{\uparrow}$ = 0.16,
  $\Gamma_{2}^{\downarrow} = 0.5$ (MeV).}
\label{fig5}
\end{figure}

\begin{figure}
\begin{minipage}{15cm}
\centerline{\psfig{figure=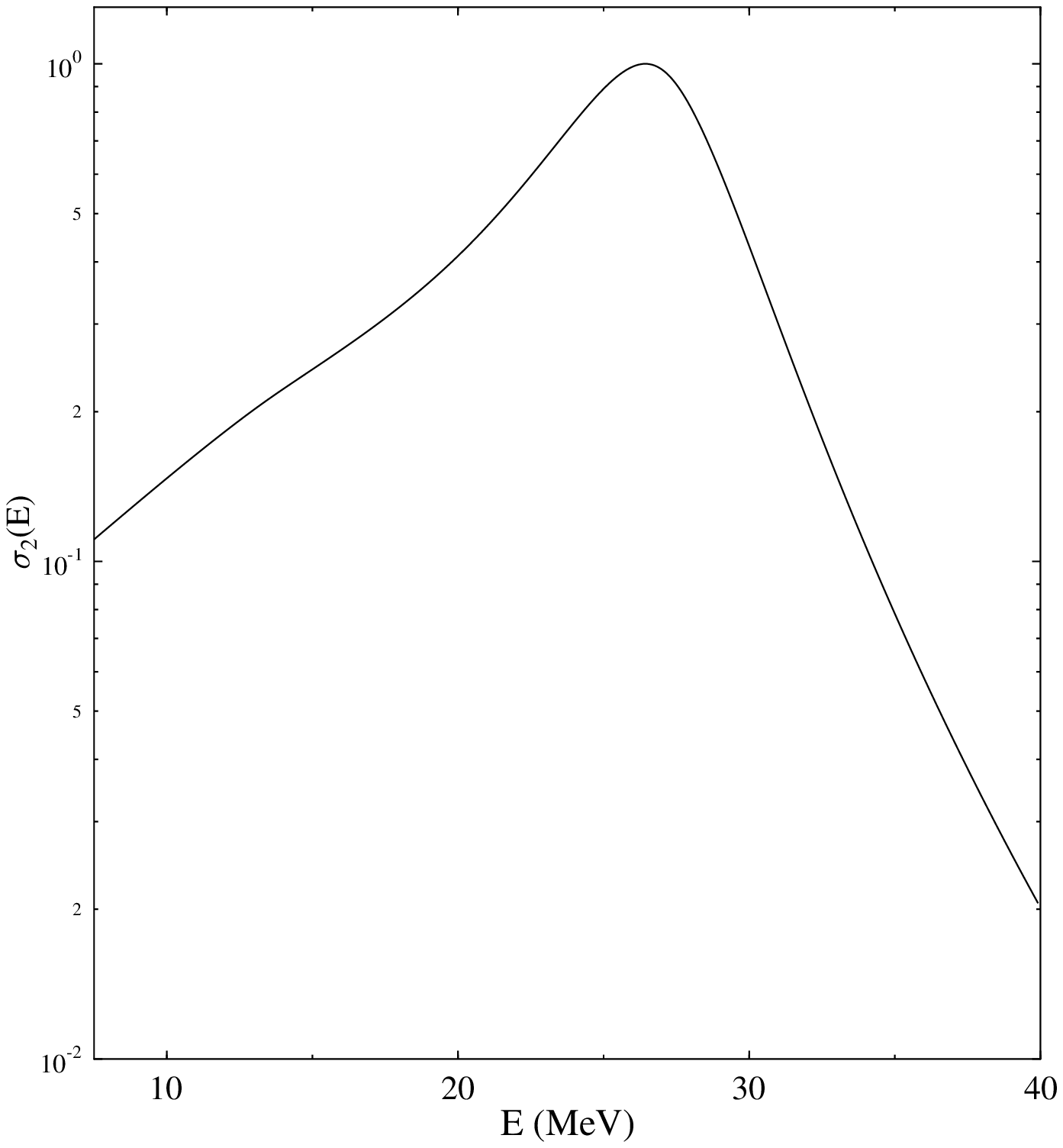,width=13cm,angle=0}}
\end{minipage}
\caption{Differential cross section for the DGDR excitation versus
excitation energy.
  Parameter values are $b_{min}$ = 15.5 fm, $E_{p}/A$ = 640, 
  $\Gamma_{10}^{\uparrow}$ = 0.11, $\Gamma_{20}^{\uparrow}$ = 0.026,
  $\Gamma_{1}^{\uparrow}$ = 0.30, $\Gamma_{2}^{\uparrow}$ = 0.16,
  $\Gamma_{10}^{\downarrow}$ = 4.0,
  $\Gamma_{20}^{\downarrow} = 0.5 \Gamma_{10}^{\downarrow}$ and
  $\Gamma_{2}^{\downarrow} = 0.5$ (MeV).}
\label{fig6}
\end{figure}

We turn to a comparison of our results with experimental data
\cite{bor96}. In keeping with experimental results \cite{sc88}, we use
122\% of the sum rule for all dipole matrix elements. Fig.~6 shows
the differential cross section for the DGDR excitation as a function 
of excitation energy. The cross section has been normalized to its
maximum value. The width of the DGDR is 6.1 MeV. Integrating the
differential cross section over excitation energy from neutron
threshold at 7.5 MeV up to 40.0 MeV (this corresponds to the
experimental situation), we find the value 410 mb. This value is
somewhat larger than that given in Ref.~\cite{bor96}, 380 $\pm$ 40 mb,
where the folding method \cite{au98} is used. Both our DGDR width and
energy--integrated cross section are consistent, however, with the
experimental values within experimental errors. In the present case,
the enhancement factor $R_{1}$ is 1.03, almost the same as shown in
Fig.~3. In Fig.~7, we plot the ratio of the energy--integrated cross
sections for the DGDR and the GDR as a function of
$\Gamma_{10}^{\downarrow}$. The ratio first increases and then
decreases as this parameter increases. At the experimental value
$\Gamma_{10}^{\downarrow} = 4.0$ MeV, the ratio is 0.0988, which again
agrees with experiment (0.11 $\pm$ 0.013 \cite{bor96}) within the
experimental error.

\begin{figure}
\begin{minipage}{15cm}
\centerline{\psfig{figure=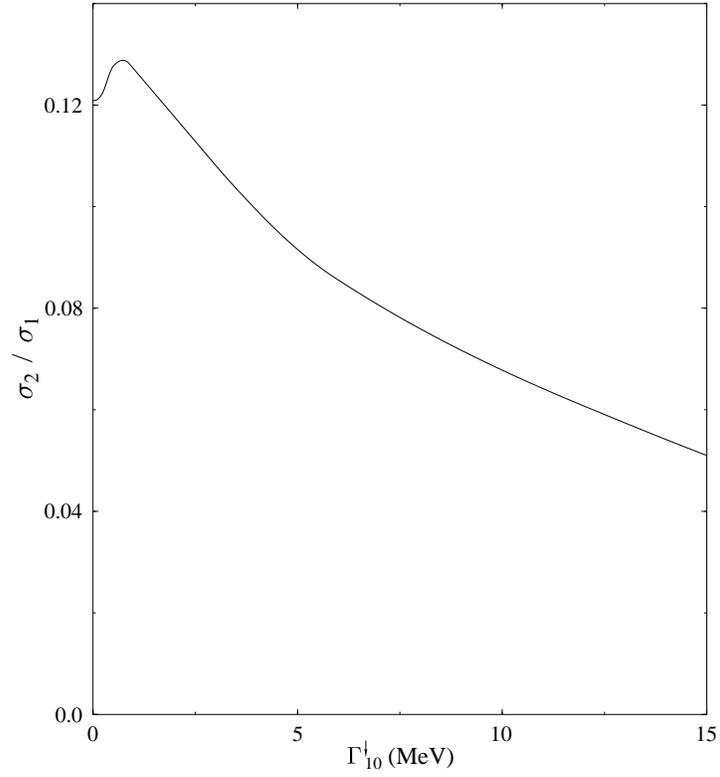,width=13cm,angle=0}}
\end{minipage}
\caption{Ratio of energy--integrated cross sections for the DGDR and
  GDR excitation versus spreading width $\Gamma_{10}^{\downarrow}$.
  Parameter values are $b_{min}$ = 15.5 fm, $E_{p}/A$ = 640,    
  $\Gamma_{10}^{\uparrow}$ = 0.11, $\Gamma_{20}^{\uparrow}$ = 0.026, 
  $\Gamma_{1}^{\uparrow}$ = 0.30, $\Gamma_{2}^{\uparrow}$ = 0.16,
  $\Gamma_{20}^{\downarrow} = 0.5 \Gamma_{10}^{\downarrow}$ and
  $\Gamma_{2}^{\downarrow} = 0.5$ (MeV).}
\label{fig7}
\end{figure}

\section{Comparison with the approach by Carlson {\it et al.}}
\label{comp}

As mentioned in Section~\ref{int}, Carlson {\it et
  al.}~\cite{car991,car992} implemented the Brink--Axel hypothesis
using a statistical model developed by Ko~\cite{ko78}. The model of
Ref.~\cite{car991,car992} differs from the one used here mainly in
terminology: The authors use a direct product representation of
Hilbert space rather than the direct sum representation of our
Eq.~(\ref{ham0}). In Refs.~\cite{car991,car992} the states in Hilbert 
space are written in the form $| n \alpha \rangle$. Here $n$ denotes
the number $n = 0,1,\ldots$ of phonons and $\alpha$ enumerates the
background states. As a consequence, the mixing of the one--phonon
state with the background states is formally described as a
de--excitation of the former, i.e., as the transition $| 1 0 \rangle
\rightarrow | 0 \alpha \rangle$. The Brink--Axel hypothesis is
automatically implemented because dipole excitation affects only the
first component $n$ in the state labelled $| n \alpha \rangle$. As far
as we can see, the physical content of both models is quite the same,
however. The main difference between both approaches lies in the
analytical evaluation of the average cross section. Indeed, in
Refs.~\cite{car991,car992} the average over the background components
labelled $\alpha$ is performed over the Green function (rather than
over the intensity) and, although intuitively reasonable, does not
involve controlled approximations. We are, in fact, not able to say
whether the statistical assumptions invoked in
Refs.~\cite{car991,car992} are the same as ours or not. In
contradistinction to this procedure, we have averaged in
Section~\ref{inte} the intensity, and we have reduced the resulting
expression by controlled approximations to the final form of
Eq.~(\ref{main}). We are sure that for realistic values of the decay
widths, the statistical error of this equation and the cross section
formulas derived from it, is less than one percent. The terminology
differs also regarding the term ``harmonic approximation''. In
Refs.~\cite{car991,car992}, this term denotes what is here called the
intraband transition.

The numerical calculations of Carlson {\it et al.}~\cite{car991,car992}
are quite different in detail from the ones performed here. This is
why we can only compare the results and not the intermediate steps.
We do this for the reaction of $^{208}$Pb on $^{208}$Pb at 640 MeV
discussed throughout this paper. For the ratio of the
energy--integrated cross section over the energy--integrated intraband
cross section (we recall that this is referred to as the harmonic
approximation cross section in Refs.~\cite{car991,car992}), Carlson
{\it et al.} find the value 120\%. This has to be compared with our
value 108\%. We conclude that the approach of Carlson {\it et
  al.}~\cite{car991,car992} constitutes a fair approximation to the
exact value.

\section{Summary}
\label{sum}

We have studied the DGDR excitation using the Brink--Axel mechanism.
The background states which couple to the one--phonon and two--phonon
states are described in terms of the Gaussian Orthogonal Ensemble of
random matrices. We use second--order time--dependent perturbation
theory and calculate analytically the ensemble--averaged cross section
for the DGDR excitation. This quantity is a function of the various
decay widths and spreading widths of the one--phonon, two--phonon, and
background states. The decay widths have been calculated from the
optical model and the exciton model. The spreading widths are taken
from experiment or used as parameters. We have numerically studied the
dependence of the cross section for excitation of the DGDR on the
various parameters of the theory. We have shown that for realistic
values of these parameters, the contribution to the cross section from
the Brink--Axel mechanism is significant. This is especially true for
the reaction $^{208}$Pb + $^{208}$Pb at projectile energy 640
MeV/nucleon. We have compared our results with those of Carlson {\it
  et al.} and have found that their work provides a fair
approximation. For sensible values of the various spreading widths, we
find that the width of the DGDR, the value of the energy--integrated
cross section, and the ratio of this quantity over the
energy--integrated cross section for single GDR excitation, all agree
with experiment within experimental errors. We take this as a strong
indication that the present approach accounts quantitatively for
the DGDR excitation. It would not be difficult to use our
Eq.~(\ref{main}) for the analysis of other data sets. Clearly, the
formalism can be extended to triple--phonon excitation.

{\bf Acknowledgment.}
We are grateful to A. Bulgac, B. Carlson, L. Canto, H. Emling,
M. Hussein, C. Lewenkopf, and J. Li for discussions, and to H. Emling
for very helpful suggestions.


\begin{thebibliography}{40}
\bibitem{sch93}R. Schmidt {\it et al.}, Phys. Rev. Lett. {\bf 70}
  (1993) 1767.
\bibitem{aum93}T. Aumann {\it et al.}, Phys. Rev. C {\bf 47} (1993)
  1728.
\bibitem{rit93}J. Ritman {\it et al.}, Phys. Rev. Lett. {\bf 70}
  (1993) 533.
\bibitem{bee93}J. R. Beene, Nucl. Phys. A {\bf 569} (1993) 163c.
\bibitem{bau86}G. Baur and C. A. Bertulani, Phys. Lett. B {\bf 174}
  (1986) 23.
\bibitem{aum95}T. Aumann, C. A. Bertulani and K. S\"ummerer,
  Phys. Rev. C {\bf 51} (1995) 416.
\bibitem{eml94}H. Emling, Prog. Part. Nucl. Phys. {\bf 33} (1994)
  729.
\bibitem{vol95}C. Volpe {\it et al.}, Nucl. Phys. A {\bf 589} (1995)
  521.
\bibitem{bor97}P. F. Bortignon and C. H. Dasso, Phys. Rev. C {\bf 56}
  (1997) 574.
\bibitem{hus99}M. S. Hussein, A. F. R. de Toledo Piza and O. K. Vorov,
  Phys. Rev. C {\bf 59} (1999) R1242.
\bibitem{car991}B. V. Carlson {\it et al.}, Ann. Phys. (N. Y.) {\bf
  276} (1999) 111.
\bibitem{car992}B. V. Carlson {\it et al.}, Phys. Rev. C {\bf 60}
  (1999) 014604.
\bibitem{bri62}D. Brink, D. Phil. thesis, Oxford University
  (unpublished), 1955; P. Axel, Phys. Rev. {\bf 126} (1962) 671.
\bibitem{ko78}C. M. Ko, Z. Phys. A {\bf 286} (1978) 405.
\bibitem{meh91}M. L. Mehta, Random Matrices, 2nd Ed., Academic Press
  (New York) 1991.
\bibitem{guh98}T. Guhr, A. M\"uller-Groeling and H. A. Weidenm\"uller,
  Phys. Rep. {\bf 299} (1998) 189.
\bibitem{ald65}K. Alder and A. Winther, Coulomb Excitation, Academic
  Press (New York) 1965.
\bibitem{ver85}J. J. M. Verbaarschot, H. A. Weidenm\"uller and
  M. R. Zirnbauer, Phys. Rep. {\bf 129} (1985) 367.
\bibitem{gu99}J. Z. Gu and H. A. Weidenm\"uller, Nucl. Phys. A {\bf
  660} (1999) 197.
\bibitem{she73}E. Sheldon and V. C. Rogers, Computer
  Phys. Commun. {\bf 6} (1973) 99.
\bibitem{man76}G. Mantzouranis, H. A. Weidenm\"uller and D. Agassi,
  Z. Phys. A {\bf 276} (1976) 145.
\bibitem{ber94}C. A. Bertulani and V. Zelevinsky, Nucl. Phys. A {\bf
  568} (1994) 931.
\bibitem{lan97}E. G. Lanza, Nucl. Phys. A {\bf 613} (1997) 445.
\bibitem{ben89}C. J. Benesh, B. C. Cook and J. P. Vary, Phys. Rev. C
  {\bf 40} (1989) 1198.
\bibitem{kox87} S. Kox {\it et al.}, Phys. Rev. C {\bf 35} (1987)
  1678.
\bibitem{ber83}G. F. Bertsch, P. F. Bortignon and R. A. Broglia,
  Rev. Mod. Phys. {\bf 55} (1983) 287.
\bibitem{her92}M. Herman, G. Reffo and H. A. Weidenm\"uller,
  Nucl. Phys. A {\bf 536} (1992) 124.
\bibitem{ber88}C. A. Bertulani and G. Baur, Phys. Rep. {\bf 163}
  (1988) 299. 
\bibitem{llo90}W. J. Llope and P. Braun-Munzinger, Phys. Rev. C {\bf
  41} (1990) 2644.
\bibitem{bor96}K. Boretzky {\it et al.}, Phys. Lett. B {\bf 384}
  (1996) 30.
\bibitem{sc88}D. Schelhaas {\it et al.}, Nucl. Phys. {\bf A 489}
  (1988) 189.
\bibitem{au98}T. Aumann, P. F. Bortignon, and H. Emling,
  Ann. Rev. Nucl. Part. Sci. {\bf 48} (1998) 351.
\end{thebibliography}
\end{document}